\documentclass[useAMS,usegraphicx,usenatbib]{mn2e}

\def\etal{{et al.\ }}
\def\eg{{\it e.g.\ }}

\def\ie{{\it i.e.\ }}
\def\cf{{\it cf.\ }}

\def\spose#1{\hbox to 0pt{#1\hss}}

\newcommand{\apprla}{\mathrel{\raise2pt\hbox{\rlap{\hbox{\lower6pt\hbox{$\sim$}}}\hbox{$<$}}}}
\newcommand{\apprga}{\mathrel{\raise2pt\hbox{\rlap{\hbox{\lower6pt\hbox{$\sim$}}}\hbox{$>$}}}}

\def\ud{{\mathrm d}}

\newcommand{\greeksym}[1]{{\usefont{U}{psy}{m}{n}#1}}

\newcommand{\uDelta}{\mbox{\greeksym{D}}}

\title[NMAGIC: Fast Parallel Implementation of a $\chi^2$-Made-To-Measure
Algorithm]{NMAGIC: Fast Parallel Implementation of a $\chi^2$-Made-To-Measure
Algorithm for Modeling Observational Data} 

\author[F. De Lorenzi, V. P. Debattista, O. Gerhard and N. Sambhus]{Flavio De Lorenzi$^{1,2}$\thanks{E-mail: lorenzi@exgal.mpe.mpg.de}, Victor P. Debattista$^{3}$\thanks{Brooks Prize Fellow},
Ortwin Gerhard$^1$, Niranjan Sambhus$^2$
\\
$^1$ Max-Planck-Institut f\"ur Ex. Physik,
Giessenbachstra{\ss}e, D-85741 Garching, Germany \\
$^2$ Astron. Institut, Universit\"at Basel, Venusstrasse 7,
Binningen, CH-4102, Switzerland \\
$^3$ Astronomy Department, University of Washington, Box
351580, Seattle WA 98195, USA}

\begin{document}   
   
\date{Accepted ---. Received ---; in original form ---}

\pagerange{\pageref{firstpage}--\pageref{lastpage}} \pubyear{----}
  
\maketitle

\label{firstpage}

\begin{abstract}   
We describe a made-to-measure algorithm for constructing $N$-particle
models of stellar systems from observational data ($\chi^2$M2M),
extending earlier ideas by Syer and Tremaine.  The algorithm properly
accounts for observational errors, is flexible, and can be applied to
various systems and geometries. We implement this algorithm in a
parallel code NMAGIC and carry out a sequence of tests to illustrate
its power and performance: (i) We reconstruct an isotropic Hernquist
model from density moments and projected kinematics and recover the
correct differential energy distribution and intrinsic kinematics.
(ii) We build a self-consistent oblate three-integral maximum rotator
model and compare how the distribution function is recovered from
integral field and slit kinematic data. (iii) We create a non-rotating
and a figure rotating triaxial stellar particle model, reproduce the
projected kinematics of the figure rotating system by a non-rotating
system of the same intrinsic shape, and illustrate the signature of
pattern rotation in this model. From these tests we comment on the
dependence of the results from $\chi^2$M2M on the initial model, the
geometry, and the amount of available data.
\end{abstract}   


\section{Introduction}   
\label{sec:introduction}   

Understanding the structure and dynamics of galaxies requires
knowledge of the total gravitational potential and the distribution of
stellar orbits. Due to projection effects the orbital structure is not
directly given by observations.  In equilibrium stellar systems, the
phase-space distribution function (DF) fully determines the state of
the galaxy. Dynamical models of observed galaxies attempt to recover
their DF and total (\ie due to visible and dark matter) gravitational
potential consistent with the observational data.  Several methods to
tackle this problem exist.  Jean's theorem (\eg \citealt{bin_tre87})
requires that the DF depends on the phase-space coordinates only
through the integrals of motion.  If these integrals can be expressed
or approximated in terms of analytic functions, one can parametrize
the DF explicitly.  This approach has been applied to spherical or
other integrable systems (\eg \citealt{dejonghe84}, \citealt{dejonghe86};
\citealt{bishop87}; \citealt{dejong_zee88}; \citealt{gerhard91};
\citealt{hunter_zee92}; \citealt{carollo_etal95}, \citealt{krona_etal00});
nearly integrable potentials where perturbation theory can be used
(\eg \citealt{saaf68}; \citealt{dehnen_ger93}, \citealt{matth_ger99}) and to
axisymmetric models assuming that the DF is a function of energy $E$
and angular momentum $L_z$ only ( \citealt{hunter_qian93};
\citealt{dehnen_ger94}; \citealt{kuijken95}; \citealt{qian_etal95};
\citealt{magorrian95}; \citealt{merritt96}).  However there is no physical
reason why the DF should only depend on the classical integrals and
most orbits in axisymmetric systems have an approximate third integral
of motion, which is not known in general \citep{ollongren62}.

\citet{schwarzschild79} developed a technique for numerically building
self-consistent models of galaxies, without explicit knowledge of the
integrals of motion.  In this method, a library of orbits is computed
and orbits are then superposed with positive definite weights to
reproduce observed photometry and kinematics.  The Schwarzschild
method has been used to model stellar systems for measurements of
global mass-to-light ratios, internal kinematics and the masses of
central supermassive black holes (\eg \citealt{rix_etal97}; 
\citealt{cretton_etal99}; \citealt{romano_kock01};
\citealt{cappellari_etal02}; \citealt{verolme_etal02};
\citealt{gebhardt_etal03}; \citealt{vdven_etal03}; \citealt{valluri_etal04},
\citealt{copin_etal04}, \citealt{thomas_etal05}). The method is well-tested, and
modern implementations are quite efficient. However, it also has some
draw-backs: symmetry assumptions are often made, and the potential
must be chosen a priori. Initial conditions for a representative orbit
library have to be carefully chosen, which becomes more complicated as
the complexity of the potential's phase space structure increases, in
terms of number of orbit families, resonances, chaotic and
semi-chaotic regions.  As a result, most Schwarzschild models in the
literature to date are axisymmetric.

Thus there is scope for exploring alternative
approaches. \citet[hereafter ST96]{ST96} invented a particle-based
algorithm for constructing models of stellar systems. This
``made-to-measure'' (M2M) method works by adjusting individually
adaptable weights of the particles as a function of time, until the
model converges to the observational data.  The first practical
application of the M2M method constructed a dynamical model of the
Milky Way's barred bulge and disk \citep{bissantz_etal04} and was able
to match the event timescale distribution of microlensing events
towards the bulge. This paper illustrates some of the promise that
lies in particle-based methods, in that it was relatively easy to
model a rapidly rotating stellar system.  However, other important
modeling aspects were not yet implemented, such as a proper treatment
of observational errors.  The purpose of the present paper is to show
how this can be done, and to describe and test our modified
$\chi^2$M2M method designed for this purpose.

The paper is organized as follows. In the Section \ref{sec:M2M} we
describe the M2M algorithm of ST96. Then in Section~\ref{sec:chi2m2m}
we extend the algorithm in order to include observational errors. We
also discuss how we include density and kinematic observables in the
same model, and describe the NMAGIC code, our parallel implementation
of the $\chi^2$M2M method. In Section \ref{sec:Targets} we present the
models we use to test this implementation, and the results of these
tests follow in Section \ref{sec:Tests}. Finally, the paper closes
with the conclusions in Section \ref{sec:Conclusions}.


\section{Syer \& Tremaine's Made-To-Measure Algorithm}
\label{sec:M2M}

The M2M algorithm is designed to build a particle model to match the
observables of some target system.  The algorithm works by varying the
individually adaptable weights of the particles moving in the global
potential until the model minimizes deviations between its observables
and those of the target.
An observable of a system characterized by a distribution function
$f(\mathbf z)$, is defined as
\begin{equation}
Y_j = \int K_j({\mathbf z})f({\mathbf z})~\ud^6z
\end{equation}
where $K_j$ is a known kernel and $\mathbf z = (\mathbf r, \mathbf v
)$ are phase-space coordinates. Examples of typical observables
include surface or volume densities and line-of-sight kinematics. The
equivalent observable of the particle model is given by
\begin{equation}
y_j(t)=\sum_{i=1}^N w_i K_j \left[ {\mathbf z}_i(t) \right],
\label{eqn:obs}
\end{equation}
where $w_i$ are the weights and ${\mathbf z}_i$ 
are the phase-space coordinates of the particles, $i=1,\cdots,N$.
In the following, we use units and normalization such that
\begin{equation}
\sum_{i=1}^N w_i = 1,
\label{eqn:norm}
\end{equation}
so that the equivalent masses of the particles are $m_i=w_i M$
with $M$ the total mass of the system.

Given a set of observables $Y_j,\ j=1,\cdots,J$, we want to construct
a system of $N$ particles $i=1,\cdots,N$ orbiting in the potential,
such that the observables of the system match those of the target
system.  The heart of the algorithm is a prescription for changing
particle weights by specifying the ``force-of-change'' (hereafter
FOC):
\begin{equation}
\frac{\ud w_i(t)}{\ud t} = -\varepsilon w_i(t) \sum_j \frac{K_j
\left[{\mathbf z}_i(t)\right]}{Z_j} \uDelta _j(t)
\label{eqn:FOC}.
\end{equation}
Here 
\begin{equation}
\uDelta _j(t) = \frac{y_j(t)}{Y_j}-1 
\label{eqn:Delta}
\end{equation}
measures the deviation between target and model observables.  The
constant $\varepsilon$ is small and positive and, to this point, the
$Z_j$ are arbitrary constants. The linear dependence of the FOC
for weight $w_i$ on $w_i$ itself ensures that the particle
weights cannot become negative, and the dependence on the kernel
$K_j$ ensures that a mismatch in observable $j$ only has influence
of the weight of particle $i$ when that particle actually 
contributes to the observable $j$. The choice of $\uDelta$ in
terms of the ratio of the model and target observables makes
the algorithm closely related to Lucy's \citeyearpar{lucy74} method, in
which one iteratively solves an integral equation for the
distribution underlying the process from observational data.

Since in typical applications the number of particles greatly exceeds
the number of independent constraints, the solutions of the set of
differential equations (\ref{eqn:FOC}) are under-determined, \ie the
observables of the particle model can remain constant, even as the
particle weights may still be changing with time.  To remove this
ill-conditioning, ST96 maximize the function
\begin{equation}
F = \mu S-\frac{1}{2}\chi^2, 
\label{eqn:SF}
\end{equation}
with
\begin{equation}
\chi^2 = \sum_j\Delta_j^2 
\label{eqn:relchi2}
\end{equation} 
and the entropy 
\begin{equation}
S = -\sum_i w_i \log (w_i/\hat{w}_i)
\label{eqn:entropy}
\end{equation}
as a profit function. The $\{\hat{w}_i\}$ are a predetermined
set of weights, the so-called priors.  Since
\begin{equation}
\mu \frac{\partial S}{\partial w_i} = -\mu (\log(w_{i}/\hat{w}_i)+1),
\label{eqn:dSdw} 
\end{equation}
if a particle weight $w_i < \hat{w}_i/e$ then equation
(\ref{eqn:dSdw}) becomes positive while it is negative when $w_i >
\hat{w}_i/e$.  Therefore the entropy term pushes the particle
weights to remain close to their priors (more specifically, close to
$\hat{w}_i/e$).
Equation~(\ref{eqn:FOC}) is now replaced by
\begin{equation}
\frac{\ud w_i(t)}{\ud t} = \varepsilon w_i(t)\left(\mu \frac{\partial
S}{\partial w_i}(t) - \sum_j \frac{K_j \left[{\mathbf
z}_i(t)\right]}{Y_j} \uDelta_j(t) \right), 
\label{eqn:modFOC}
\end{equation} 
with $Z_j$ now fixed to $Y_j$ by the requirement that
equation~(\ref{eqn:SF}) will be maximized, as discussed in Section
\ref{sec:chi2m2m}. The constant $\mu$ governs the relative importance
of the entropy term in equation (\ref{eqn:modFOC}): When $\mu$ is
large the $\{w_i\}$ will remain close to their priors
$\{\hat{w}_i\}$. In the following, we will generally set
$\hat{w}_i=w_0=1/N$; i.e., the particle distribution follows the
initial model, but this is not necessary.

To reduce temporal fluctuations, ST96 introduced temporal smoothing by
substituting $\Delta_j(t)$ in Equations (\ref{eqn:relchi2}) and
(\ref{eqn:modFOC}) with
\begin{equation}
\widetilde{\uDelta} _j(t) = \alpha \int_0^{\infty}
\Delta_j(t-\tau)e^{-\alpha \tau} \ud \tau,
\label{eqn:Sts},
\end{equation}
which can be expressed in the form of the differential equation
\begin{equation}
\frac{\ud \widetilde{\uDelta} }{\ud t} = \alpha
\left(\uDelta-\widetilde{\uDelta}\right) \label{ts}.
\end{equation}
The smoothing time is $1/\alpha$. The temporal smoothing suppresses
fluctuations in the model observables and hence in the FOC correction
of the particle weights -- in the computation of these quantities the
effective number of particles is increased as each particle is
effectively smeared backwards in time along its orbit.  The smoothing
time should satisfy $2 \epsilon < \alpha$ to avoid excessive temporal
smoothing\footnote{This corrects the typo in equation~(19) of ST96.},
which slows down convergence.


\section{$\chi^2$-based Made-to-Measure Algorithm to Model Observational Data}
\label{sec:chi2m2m}

The M2M algorithm as originally formulated by ST96 is well adapted to
modeling density fields (\eg \citealt{bissantz_etal04}).  It is not,
however, well suited to mixed observables such as densities and
kinematics, where the various ratios of model to target observable can
take widely different values, or to problems where
observables can become zero, when $\uDelta$ diverges.  Moreover, the
$\chi^2$ defined as in equations~(\ref{eqn:relchi2},\ref{eqn:Delta})
is not the usual one, but is given by the relative deviations between
model and data. Thus extremizing $F$ (equation~\ref{eqn:SF}) with this
$\chi^2$ does not produce the best model given the observed
data.  We have therefore modified the M2M method as described in this
section.

We begin by considering observational errors.  We do this by replacing
equation~(\ref{eqn:Delta}) by
\begin{equation}
\uDelta_j(t) = \frac{y_j-Y_j}{\sigma(Y_j)},
\label{eqn:myDelta}
\end{equation}
where $\sigma(Y_j)$ in the denominator is the error in the target
observable.  With this definition of $\uDelta_j$
equation~(\ref{eqn:relchi2}) now measures the usual absolute $\chi^2$.
As a result of this, if we now maximize the function of
equation~\ref{eqn:SF} with respect to the $w_i$'s we obtain the
condition
\begin{equation}
\mu \frac{\partial S}{\partial w_i} - \sum_j
\frac{K_{ji}}{\sigma(Y_j)}\uDelta_j = 0. 
\label{eqn:condFmax}
\end{equation} 
If we replace the FOC equation~(\ref{eqn:modFOC}) by
\begin{equation}
\frac{\ud w_i(t)}{\ud t} = \varepsilon w_i(t)\left(\mu \frac{\partial
S}{\partial w_i} - \sum_j \frac{K_j \left[{\mathbf
z}_i(t)\right]}{\sigma(Y_j)} \uDelta_j(t) \right) 
\label{eqn:myFOC}
\end{equation} 
then the particle weights will have converged once $F$ is maximized
with respect to all $w_i$, i.e., once the different terms in 
the bracket balance. For large $\mu$, the solutions
of eqn.~\ref{eqn:myFOC} will have smooth weight distributions at the
expense of a compromise in matching $\chi^2$.

In the absence of the entropy term,
the solutions of eqs.~\ref{eqn:myFOC} near convergence can be
characterized by an argument closely similar to that used by ST96 to
study the solutions of their eqs.~(4). For small $\varepsilon$, the weights
$w_i(t)$ change only over many orbits, so we can orbit-average 
over periods $t_{\rm orb} \ll \tau \ll t_{\rm orb}/\varepsilon$ and
write the equations for the orbit-averaged $<\uDelta_j>$ as
\begin{equation}
\frac{\ud <\uDelta_j(t)>}{\ud t} = 
         \varepsilon {\cal A}_{jk} <\uDelta_k(t)>,
\label{eqn:orbav}
\end{equation}
where the matrix {\cal A} has components
\begin{equation}
{\cal A}_{jk}=\Sigma_i \frac{<K_{ji}><K_{ki}>}{\sigma_j \sigma_k} w_i^0,
\end{equation}
and we have replaced $w_i(t)$ by the constant $w_i^0$, because near
convergence the dominant time-dependence is in $<\uDelta_k>$ rather
than $w_i$.  The matrix {\cal A} is symmetric by construction and
positive definite, i.e., ${\sl x}^t \cdot {\cal A} \cdot {\sl x} > 0$
for all vectors {\sl x}; so all its eigenvalues are real and positive.
The solutions to eqs.~\ref{eqn:orbav} then converge exponentially to
$<\uDelta_j(t)> =0$. As for eqs.~(4) of ST96, this argument suggests
that if $\varepsilon$ is sufficiently small and we start close to the
correct final solution, then the model observables converge to their
correct final values on $O(\varepsilon^{-1})$ orbital periods.

Substituting $\uDelta_j$ in equation~(\ref{eqn:Sts}) leads to
\begin{equation}
\widetilde{\uDelta}_j(t) = \frac{\widetilde{y}_j(t)-Y_j}{\sigma(Y_j)},
\end{equation}
which allows us to temporally smooth model observables directly
\begin{equation}
\widetilde{y}_j(t)= \alpha \int_0^{\infty} y_j(t-\tau)e^{-\alpha \tau}
\ud \tau 
\label{eqn:myts}.
\end{equation}
In practice, $\widetilde{y}_j$ can be computed using the equivalent
differential equation, in the same manner as before.

Since the uncertainty in any observable presumably never becomes zero,
the $\uDelta_j$ in equation~(\ref{eqn:myDelta}) remain well-defined
even when the observables themselves take zero values.  However, if
the data entering $\chi^2$ have widely different relative errors, the
FOC equation may be dominated by only a few of the $\uDelta_j$. This
can slow down convergence of the other observables and thus lead to
noisy final models.
Also, notice that the cost of deriving the FOC from minimizing
$\chi^2$ is that equation~(\ref{eqn:SF}) is maximized only if the
observables are exactly of the form given by equation~(\ref{eqn:obs}),
\ie the kernel $K_{ij}$ may depend on the particle's phase-space
coordinates but must not depend on its weight $w_i$.
 
We adopt the convention throughout this paper in which the positive
$x$-axis points in the direction of the observer, so that a particle
with velocity $v_x < 0$ will be moving away from the observer.

Our implementation of the $\chi^2$M2M algorithm models volume
luminosity densities (equivalent to luminous mass densities for
constant mass-to-light ratio), and line-of-sight velocities. As in the
Schwarzschild method, dark matter, which generally has a different
spatial distribution from the stars, can be included as an external
potential, to be added to the potential from the luminous particles.
The form of the dark matter potential can be guided by cosmological
simulations, or also include information from gas velocities and
other data. 


\subsection{Densities}

For modeling the target distribution of stars one can use as
observables the surface density or space density in various grids, or
also some functional representations such as, \eg, isophote fits,
multi-Gaussian expansions, etc. In this paper we have chosen to model
a spherical harmonics expansion of the three-dimensional density,
where we expand the density in surface harmonics computed on a 1-D
radial mesh of radii $r_k$.  The expansion coefficients, $A_{lm}$ are
computed based on a cloud-in-cell scheme. The function
\[ \gamma_k^{CIC}(r) = \left \{ \begin{array}{ll}
\frac{r-r_{k-1}}{r_k-r_{k-1}} & \mbox{if $r \in [r_{k-1},r_k)$} \\
\frac{r_{k+1}-r}{r_{k+1}-r_k} & \mbox{if $r \in [r_k,r_{k+1}]$} \\
0 & \mbox{otherwise,}
	\end{array}
\right. \]
gives the fractional contribution of the weight $w$ of a particle
at radius $r$ to shell~$k$.  The model observable is then the mass on
each shell $k$,
\begin{equation}
m_k = M \sum_i w_i \gamma_k^{CIC}(r_i) 
   \equiv M \sum_i w_i \gamma_{ki}^{CIC}.
\end{equation}
Comparing with equation~(\ref{eqn:obs}), we recognize the kernel for
this observable as
\begin{equation}
K_{ki} = M \gamma_{ki}^{CIC}.
\end{equation}
Thus the FOC on a particle is computed by linear interpolation of
the contributions from the adjacent shells.  From
equation~(\ref{eqn:myDelta}), we obtain
\begin{equation}
\uDelta_k[m] = \frac{m_k-M_k}{\sigma(M_k)}
\end{equation} 
where $M_k$ is the target mass on shell $k$ and $\sigma(M_k)$ 
its uncertainty.

The spherical harmonic coefficients for the particle model
with $l>0$ are computed via
\begin{equation}
a_{lm,k} = M \sum_i \gamma_{ki}^{CIC} Y_l^m(\theta_i,\varphi_i) w_i.
\label{eqn:alm}
\end{equation}
Now the kernel is given by
\begin{equation}
K_{{\mathbf j}i} = M \gamma_{ki}^{CIC} Y_l^m(\theta_i,\varphi_i), \,\,\, 
   {\mathbf j} =\{lm,k\},
\end{equation}
and depends on the spherical harmonics; the same variation also holds
therefore for the FOC.  From equation~(\ref{eqn:myDelta}), we obtain
\begin{equation}
\uDelta_{\mathbf j}[a_{lm}] = \frac{a_{lm,k}-A_{lm,k}}{\sigma(A_{lm,k})},
 \,\,\, {\mathbf j}=\{lm,k\},
\end{equation}
with $A_{lm,k}$ as the target moments and $\sigma(A_{lm,k})$ as their
errors.  $a_{00,k}$ and $A_{00,k}$ are of course related to the mass
on shell $k$ via the relation $\sqrt{4\pi}a_{00,k} = m_k$, etc., but
we will use the masses on shells $m_k$, $M_k$ as observables in the
following.

\subsection{Kinematics}

Unlike for the density observables, we use kinematic observables
computed in the plane of the sky to compare with the target model.
Since kinematic data can either come from restricted spatial regions
(\eg slit spectra) or from integral fields, we do not specify any
special geometry for computing these observables.

The shape of the line-of-sight velocity distribution (LOSVD) can be
expressed in a truncated Gauss-Hermite series with coefficients
$h_n,\;n=1,\cdots,n_{\rm max}$ (\citealt{gerhard93},
\citealt{vdmarel_franx93}). Since the kernel in
equation~(\ref{eqn:myFOC}) cannot depend on masses, this puts some
constraints on which observables can be used in the FOC.  For
kinematics, suitable observables are the {\it mass-weighted}
Gauss-Hermite coefficients, which we use as follows.  Particle weights
are assigned to a spatial cell, $\mathcal{C}_p$, of the kinematic
observable under consideration using the selection function
\[ \delta_{pi} = \left \{ \begin{array}{ll}
1 & \mbox{if $(y_i,z_i) \in \mathcal{C}_p$} \\
0 & \mbox{otherwise.}
	\end{array}
\right. \]
This selection function can be replaced appropriately if seeing
conditions need to be taken into account.  In our present application
this is not necessary.  The mass-weighted kinematic moments
are computed as
\begin{equation}
b_{n,p} \equiv m_p\, h_{n,p} = 2\sqrt{\pi}M\sum_i \delta_{pi} u_n(\nu_{pi}) w_i, 
\end{equation}
\begin{equation}
\nu_{pi}=\left(v_{x,i}-V_p\right)/\sigma_p,
\end{equation}
and where $m_p$ is the mass in cell $\mathcal{C}_p$, and
the dimensionless Gauss-Hermite functions \citep{gerhard93}
\begin{equation}
u_n(\nu) = \left(2^{n+1}\pi n!\right)^{-1/2}H_n(\nu)\exp \left(-\nu^2/2\right).
\end{equation} 
$H_n$ are the standard Hermite polynomials.  For the mass-weighted
higher order moments we obtain the kernel
\begin{equation}
K_{{\mathbf j}i} = 2 \sqrt{\pi} M \delta_{pi} u_n(\nu_{pi}),  \,\,\, 
  {\mathbf j}=\{n,p\}.
\end{equation} 
and as usual
\begin{equation}
\uDelta_{\mathbf j} [m\, h_n] = \frac{b_{n,p} - B_{n,p}}{\sigma(B_{n,p})},  
  \,\,\, {\mathbf j}=\{n,p\}.
\end{equation}
The velocity $V_p$ and dispersion $\sigma_p$ are not free parameters;
rather we set $V_p$ and $\sigma_p$ to the mean line-of-sight velocity
and velocity dispersion obtained from the best fitting Gaussian to the
observed (target) LOSVD. This implies $ B_{1,p}\equiv (m_p\,
h_{1,p})_{\rm target} = B_{2,p} \equiv (m_p\, h_{2,p})_{\rm target} =
0$ for the first and second order mass-weighted target Gauss-Hermite
coefficients. If the model $b_{1,p}$ and $b_{2,p}$ both converge to
zero, then the LOSVD of the particle model automatically has the
correct mean line-of-sight velocity and velocity dispersion.  For
describing the higher-order structure of the LOSVD we include terms
$m\,h_n$ $(n=1,\cdots,4)$ in the test modeling described below.

\subsection{Implementation: the NMAGIC parallel code}

\begin{figure*}
\includegraphics[width=0.8\hsize,angle=0.0]{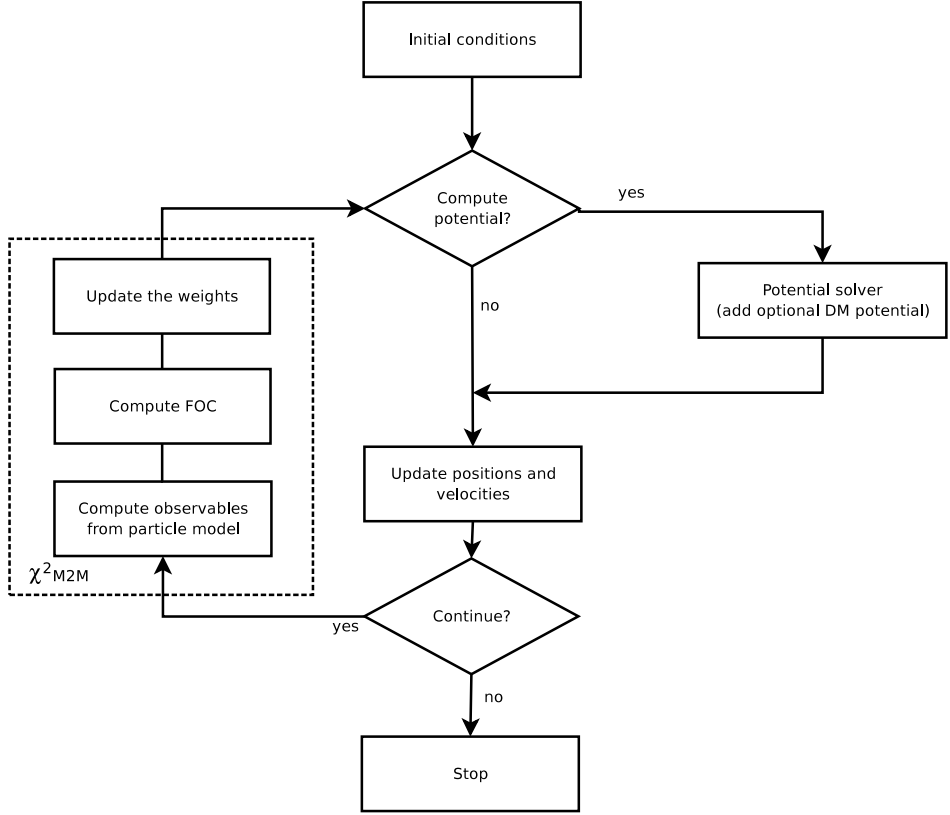}
\caption[]{
A high level flowchart describing NMAGIC.  The main $\chi^2$M2M
algorithm is contained in the dashed block, the remainder is an
optional potential solver and code for moving the particles, both of
which are exchangeable. In our tests, $\chi^2$M2M is generally
applied only after a number of position/velocity updates.
}
\label{fig:flowchart} 
\end{figure*}

The routine for updating the particle weights includes three main
steps: First, all the observables used in the modeling process are
computed as described above. Then we change the particle weights in
accordance with equation~(\ref{eqn:myFOC}) by
\begin{equation}
w_{i,t+\delta t} = w_{i,t}+\varepsilon w_{i,t}\left(\mu \frac{\partial
S}{\partial w_i} - \sum_j \frac{K_j \left[{\mathbf
z}_{i,t}\right]}{\sigma(Y_j)} \widetilde{\uDelta}_{j,t} \right) \delta t,
\label{eqn:discFOC}
\end{equation}
with
\begin{equation}
\widetilde{\uDelta}_{j,t} = \frac{\widetilde{y}_{j,t}-Y_j}{\sigma(Y_j)}.
\end{equation}
Finally, we update the temporally smoothed observables
as follows:
\begin{equation}
\widetilde{y}_{j,t+1} = \widetilde{y}_{j,t} + \alpha
(y_{j,t}-\widetilde{y}_{j,t})\delta t.
\end{equation}
Here $\delta t$ is the time between successive $\chi^2$M2M steps.  All
the differential equations here are ordinary differential equations of
the form $\ud y_i(t) / \ud t =f_i(t,y_1,\cdots,y_N)$, and
the $y_{i,n}$'s in our case are the particle weights $w_i$ or
time-smoothed observables $y_i$ at $t_n$. We integrate them
using a simple Euler method $y_{i,n+1} = y_{i,n}+h\;f(t_n,y_{i,n})$
with $t_{n+1}=t_n+h$ and time step $h=\delta t$.  We could replace the
Euler method by the second-order Runge-Kutta method (\cf \citealt{press_etal92}),
which is more expensive and requires more memory.  Since we are not
interested in the details of how the weights converge, but only in the
final converged system, a simple Euler method suffices for our
purposes.  We write $\varepsilon$ in equation~(\ref{eqn:discFOC}) as
$\varepsilon = \varepsilon' \varepsilon''$ with $\varepsilon''=
10/\max_{i,j}\{ K_{ji}\, \widetilde{\uDelta}_j / \sigma(Y_j) \}$.
Thus $\varepsilon''$ times the last term in equation~(\ref{eqn:myFOC})
is of order unity and we choose $2 \varepsilon'<\alpha$ to avoid
excessive temporal smoothing.

The NMAGIC ({\bf N}-particle {\bf M}ade-to-measure {\bf A}l{\bf
G}orithm m{\bf I}nimizing {\bf C}hi squared) correction routine can be
combined with a standard $N$-body code including a potential solver
and time integrator, or a fixed-potential routine and integrator when
the target is to be modeled in a given gravitational potential. This
last case is most similar to the Schwarzschild method. In most of the
tests below, we use a fixed potential expanded in spherical harmonics.

However, in test E we allow the potential to vary, as we evolve from
one triaxial model to another. For advancing the particles we use a
standard leap frog time integrator with fixed time-step.  The
time-step value chosen leads to fluctuations of energy and angular
momentum with amplitudes $5\times 10^{-6}$ and $2\times 10^{-5}$
around their initial values, without systematic drift, over 80 half
mass dynamical times in the fixed potential case.

For test E, which models a triaxial system, a simple spherical
harmonic expansion suffices for solving for the potential.  We follow
the method described by \citet{sellwood03}: we tabulate coefficients
of a spherical harmonic expansion of the density on a 1-D radial grid
but retain the exact angular dependence up to some adopted $l_{\rm max}$,
the maximum order of the spherical harmonic expansion.  We include
terms up to $l_{\rm max} = 4$ in this experiment.  Particles are binned on
the radial grid using the scheme described by \citet{sellwood03}.
This then gives the forces on the grid, from which we interpolate back
to a particle's position for the gravitational forces.  Test E
involves a cuspy model; in order to properly resolve this we use a
radial grid at radii $r_\xi = e^{\gamma \xi} -1$ with $\gamma =
\ln(r_{\rm max} + 1)/\xi_{\rm max}$; we use $\xi_{\rm max} = 301$ for 301
shells and $r_{\rm max} = 40$.

NMAGIC is written in Fortran 90 and parallelized with the MPI library.
We distribute the $N$ particles as nearly evenly as possible over
$N_p$ processors.  Parallelizing in only the observables would not
scale well with large $N_p$, because of the different nature of the
observables, and would require a large memory on each processor when
$N$ is large.  In Figure \ref{fig:flowchart} we present a high level
flowchart of the operational logic of NMAGIC.

\begin{figure}
\includegraphics[angle=-90.0,width=0.95\hsize]{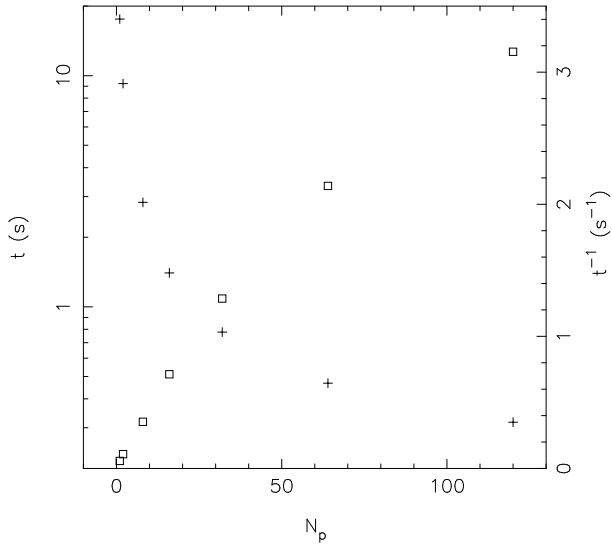}
\caption[]{ The performance of our implementation of $\chi^2$M2M.  We
used $1.8\times 10^6$ particles without potential calculations or
particle motion.  On the left hand axis we label time per step
required, with the corresponding data indicated by plus symbols, while
on the right hand axis we label steps per unit time, with the
corresponding data now shown by open squares.  Note that the scale is
logarithmic on the left and linear on the right.  The fraction of
sequential code, $f$, from these data was computed at $\sim 1\%$.  }
\label{fig:scaling}
\end{figure}

In order to test the scaling with $N_p$ of NMAGIC we considered $N =
1.8 \times 10^6$ and $N_o = 816$ observables (640 density and 176
kinematic) with $N_p$ varying from 1 to 120.  These values of $N_p$
and $N_o$ are adequate for the experiments presented here and are used
in test C of Table \ref{table:Tests}.  Since we are only interested in
the scaling of the $\chi^2$M2M parallelization with $N_p$, we only
execute the $\chi^2$M2M 50 times, without recomputing the potential or
advancing particles.  In Figure \ref{fig:scaling} we present these
scaling results as time per step (left hand axis, plus symbols) and
steps per unit time (right hand axis, open squares) as functions of
$N_p$.  We generally find that our implementation of $\chi^2$M2M
scales very well with $N_p$.  Defining the speedup $S(N_p,N)$ as
\begin{equation} 
S(N_p,N) = \frac{T(1,N)}{T(N_p,N)} 
\end{equation} 
where $T(N_p,N)$ is the time for computing $N$ particles on $N_p$
processors, we fit a standard Amdahl's law \citep{amdahl67}
\begin{equation}
S(N_p,N) = \frac{1}{f+(1-f)/N_p},
\label{eqn:amdahl}
\end{equation}
in order to determine the fraction of sequential code, $f$.  We
obtained that $f \simeq 0.010$, i.e. the sequential part of the code
accounts for only $1\%$.

\section{Target Models and Their Observables}
\label{sec:Targets}

We will test the NMAGIC code on spherical, axisymmetric, and triaxial
target models. The spherical target is a particle model constructed
from the analytic density and distribution function of an isotropic
Hernquist sphere.  As oblate target we take a maximally rotating
three-integral model. Finally, we construct both a stationary and a
rotating triaxial target system. We use the NMAGIC code itself to
generate a dynamical equilibrium structure for these models. It will
be seen that the $\chi^2$M2M method provides a very useful means to
set up dynamical equilibrium models of galaxies for which no analytic
distribution functions are known, in order to study the properties of
such systems.

In the following subsections, we describe in turn each of these
targets and their construction. We determine the target observables
obtained from these models, and describe how we obtain errors for
these observables. These will be needed in Section \ref{sec:Tests}
where we present the results of building $\chi^2$M2M models to match
these targets.  The reader who is mainly interested in these tests of
NMAGIC can in a first reading directly go to that section.

\subsection{Spherical Target}
\label{ssec:sphModels}

Our first target is a spherical isotropic Hernquist \citep{hernquist90}
model, which we will refer to as target SIH.  Its density and
potential are given by
\begin{equation}
\varrho (r) = \frac{a M}{2 \pi r(r+a)^3}, \qquad
\varphi (r) = -\frac{G M}{r+a},
\label{eqn:HernDens}
\end{equation} 
where $a$ is the scale length, $M$ is the total mass, and
$G$ is the gravitational constant.  The projected effective
(half-mass) radius equals $R_{\rm eff}\approx 1.8153a$. We use units
such that $M = a = 1$.  The target mass $M_k$ on shell $r_k$ is given
by the sum of the contributions of the adjacent shells,
\begin{equation}
M_k = 4 \pi \int \varrho(r) \gamma^{CIC}_k(r) r^2\ud r.
\label{eqn:HernM}
\end{equation}
The innermost (outermost) shell is an exception because only the layer
immediately exterior (interior) contributes.

\begin{figure}
\includegraphics[angle=-90.0,width=0.9\hsize]{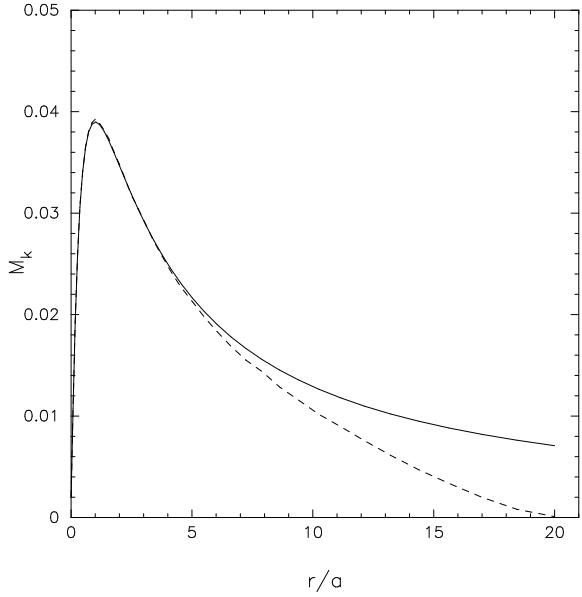}
\vskip0.5truecm
\caption[]{The mass in shells profile computed from
equation~(\ref{eqn:HernDens}) is shown as the solid line, whereas the
dashed line illustrates the mass profile computed from a spherical
Hernquist particle model generated from a truncated DF. }
\label{fig:MMp}
\end{figure}

We construct SIH models on a radial grid with 40 shells,
quasi-logarithmically spaced in radius with inner and outer boundaries
at $r_{\rm min}=5 \times 10^{-4}$ and $r_{\rm max}=20$. The distribution
function is truncated at $E_{\rm max}\equiv\phi(r_{\rm max})$.  At
that truncation, the mass included is
\begin{equation}
M_{\rm trun}=\int_{E_{\rm min}=\varphi(0)}^{E_{\rm max}=
\varphi(r_{\rm max})}\frac{dM}{dE}dE,
\end{equation}
with $(dM/dE)$ the differential energy distribution (\eg
\citealt{bin_tre87}) and thus $M_{\rm trun}=0.86$.  Figure \ref{fig:MMp}
compares the mass on shells (hereafter ``mass profile'') $M_P(r_k)$
for a particle realization of this truncated distribution function
(constructed using the method described in \citealt{victor_sell00}),
with the $M_k$ from the Hernquist density profile as in
equation~(\ref{eqn:HernM}).  For small radii the mass profiles match
but for larger radii $M_P(r_k)$ is significantly smaller than $M_k$
due to the finite extent of the particle realization, consisting only
of particles with $E<E_{\rm max}=\varphi(r_{\rm max})$. Using $M_k$ as
target observables would increase the mass of particles on the outer
(near) circular orbits and would therefore increase the tangential
velocity dispersion. We will thus use the $M_P(r_k)$ as targets and
omit the subscript $P$ in the following. We also include zero-valued
higher order mass moments to enforce sphericity.

We assume Poisson errors for the radial mass: $\sigma(M_j)=\sqrt{M_j
M_{\rm trun}/N}$ where N is the total number of particles used in the
particle model. For the errors in the higher order mass moments, we
use Monte-Carlo experiments in which we generate particle
realizations of the density field of the target model using $1.8
\times 10^6$ particles, which is the same number as in the
$\chi^2$M2M models.

Kinematics of the target can be computed from a DF.  We use the
isotropic DF (\citealt{hernquist90}, \citealt{carollo_etal95})
\begin{eqnarray}
f(E) & \propto &
 \frac{1}{(1-q^2)^{5/2}}\big(3\arcsin(q)+q(1-q^2)^{1/2} \nonumber \\
 & &\times (1-2q^2)(8q^4-8q^2-3)\big)
 \label{eqn:HernquistDF}
 \end{eqnarray}
with $q = \sqrt{-aE/GM}$, and $E$ is the energy.  We determine
kinematic observables of the target on a projected radial grid with 30
shells, quasi-logarithmically spaced in radius and bounded by $R_{\rm
min}=0$ and $R_{\rm max}=10=5.51R_{\rm eff}$.  On the shell midpoints
we compute the $h_2$ and $h_4$ moments of the isotropic Hernquist
model from the DF of equation~(\ref{eqn:HernquistDF}).  We will use
integral field-like kinematic data to recover the spherical targets in
Section~\ref{sec:Tests}. More precisely, we multiply the $h_{2,k}$ and
$h_{4,k}$ moments by the projected mass of the truncated SIH model
within each radial grid shell to obtain the mass-weighted higher order
moments $M_k\,h_{2,k}$ and $M_k\,h_{4,k}$, which we use as the target
observables.  While this procedure is not perfectly self-consistent,
because the moments are from the infinite extent analytic DF while the
mass is from the truncated DF, the differences are very small.  The
main advantage of doing this is that it allows us to compute the
uncertainties in these kinematic observables, which we assume
$\sigma(M_k h_{n,k})=\sigma(h_n) M_c\sqrt{M_k/M_c}$ with $\sigma(h_n)
= 0.005$, $M_k$ the target mass in shell $k$, and $M_c$ the mass in the
central grid shell.

\subsection{An Oblate Three-Integral Target made with NMAGIC}
\label{ssec:obModels}

\begin{figure}
\includegraphics[angle=-90.0,width=0.9\hsize]{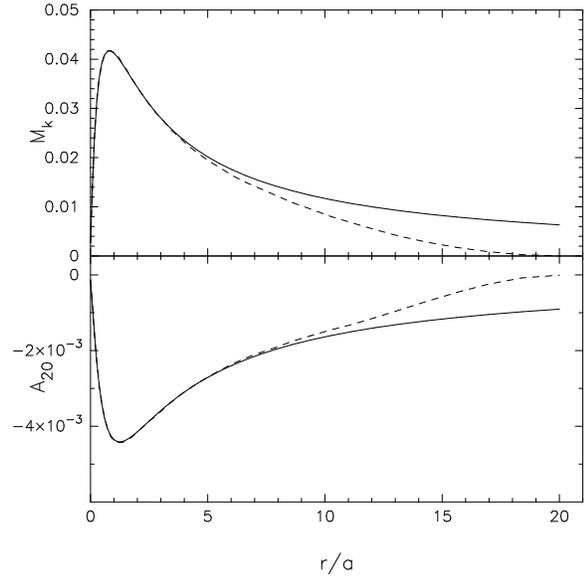}
\vskip0.5truecm
\caption[]{The upper panel shows the mass profile computed from
  equation~(\ref{eqn:obldens}) for $q=0.6$ (solid line) and from a
  Hernquist particle model made from a DF and squeezed along the
  $z$-axis (dashed line). The lower panel is the same but for
  $A_{20}$.}
\label{fig:AAp}
\end{figure}

\noindent
Our oblate target model has density
\begin{equation}
\varrho (m) = \frac{a M}{2 \pi q m(m+a)^3}
\label{eqn:obldens}
\end{equation}
where $M$ and $a$ are total mass and scale radius, and $m^2 =
R^2+(z/q)^2$ with $q$ being the flattening. This density belongs to
the family of flattened $\gamma$ models \citep{dehnen_ger94}, with
$\gamma=1$.  We compute the gravitational potential from (\cf
\citealt{bin_tre87}, section 2.3)
\begin{equation}
\varphi(R,z) = -\frac{GM}{2a}\int_0^{\infty}
\frac{\tilde{\psi}(\tilde{m})\ud \tau}{(1+\tau)\sqrt{\tau+q^2}}
\label{eqn:oblpot}
\end{equation}
with
\begin{equation}
\tilde{m}=\sqrt{\frac{R^2}{\tau+1}+\frac{z^2}{\tau+q^2}},
\end{equation}
\begin{equation}
\tilde{\psi}(m)=1-\frac{m^2+2am}{(m+a)^2}.
\end{equation}
by numerical integration, and tabulate it using a coarse and a fine
linear grid in the meridional ($R-z$) plane. The coarse grid extends to
$R=z=30a$ with $500\times 500$ grid points. To increase the resolution
at small $R$ and $z$ we replace the $20\times 20$ ``innermost'' grid
cells at $(R,z)=(0,0)$ to $(1.2a,1.2a)$ by a finer grid also
consisting of $500\times 500$ grid points.

\begin{figure}
\includegraphics[angle=-90.0,width=0.9\hsize]{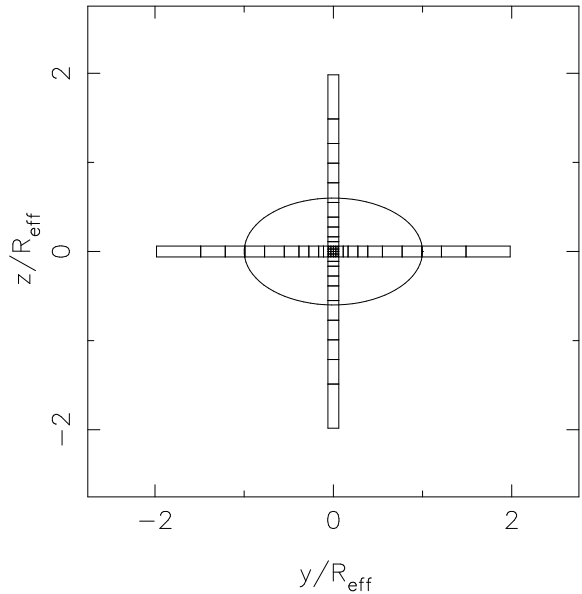}
\vskip0.5truecm
\caption[]{Kinematic major and minor axis slits for the oblate models,
  with the cells along each slit indicated. The ellipse corresponds to
  one $R_{\rm eff}$ of the equivalent Hernquist model squeezed by $0.6$ in
  the $z$ direction.}
\label{fig:slits}
\end{figure}

In our experiments, we view the model edge-on along the $x$-axis as
line-of-sight. Our targets are the mass moments $A_{lm,k}$ of the
three-dimensional density $\rho$, and -- for these oblate models --
the kinematic moments $m\;h_n$, $n = 1, ..., 4$.  We define an
effective radius $R_{\rm eff} \approx 1.8153a$ which is equal to that
of the spherical Hernquist model. We set $M = a = 1$ and $q = 0.6$.
The target mass moments $A_{lm,k}$ on shell $r_k$ are given by the sum
of the contributions of the adjacent shells and are computed through
\begin{equation}
A_{lm,k} = \int Y_{lm}(\theta,\phi)\varrho({\mathbf
x}) \gamma^{CIC}_k(r) \ud^3x.
\end{equation}
The innermost (outermost) shell is an exception because only the layer
immediately exterior (interior) contributes. The setup of the radial
grid is identical to that used for the spherical model and for our
tests below we use $M_k$, $A_{20,k}$, $A_{22,k}$, $\cdots$, $A_{66,k}$.

Figure \ref{fig:AAp} compares $M_k$ and $A_{20,k}$ 
computed from
equation (\ref{eqn:obldens}) with $M_P(r_k)$ and $A_{P,20}(r_k)$
obtained from a spherical Hernquist particle realization built from a
DF and squeezed along the $z$-axis by $q=0.6$.  As in Figure
\ref{fig:MMp}, $M_P(r_k)$ and $A_{P,20}(r_k)$ match $M_k$ and
$A_{20,k}$ within $r \apprla 5a$ but then approach zero at larger
radii towards $r_{\rm max}$.  This difference is again due to the
finite extent of the particle model. Below we therefore use the radial
mass profile $M_P$ and the higher mass moments $A_{P,lm}$ as targets,
and again we omit the subscript $P$ in the following.

We assume errors in the target mass profile $\sigma(M_j)$ as for the
spherical model.  For the errors in the higher order mass moments, we
use Monte-Carlo experiments in which particle realizations of the
density field of the target model are generated using $5 \times
10^5$ particles, which is the same number as in the $\chi^2$M2M
models.

In our oblate models we attempt to recover the target system from both
slit and integral field kinematic data.  Thus as kinematic target
observables we use the projected mass-weighted Gauss-Hermite moments
along the major and minor axes in Test C, and on a grid of $30\times
20$ points covering positions on the sky in $[-3.6,3.6]\times
[-1.8,1.8]$ in Test D. A schematic representation of the slit setup is
shown in Figure \ref{fig:slits}. The slits extend out to about
$2R_{\rm eff}\simeq 3.6$.

The target kinematics are determined from a $4\times 10^6$ particle
representation of a maximally rotating three-integral model for the
density distribution of equation~(\ref{eqn:obldens}) with
$q=0.6$. This is constructed by first evolving an isotropic spherical
Hernquist model to the desired shape, using $\chi^2$M2M, and then
switching the in-plane velocity vectors of all particles with positive
angular momentum $J_z$ to negative $J_z$, leading to a DF which is
still a valid solution of the Boltzmann equation \citep{lyndenbell60}.
For each slit or integral field cell $p$ we obtain the mass in that
cell $M_p$ and the mass-weighted Gauss-Hermite moments
$M_p\,h_{1,p},\cdots,\,M_p\,h_{4,p}$.  We assume errors for the
mass-weighted Gauss-Hermite coefficients as for the spherical model:
$\sigma(M_p h_{n,p}) = \sigma(h_n)M_c \sqrt{M_p/M_c}$, where $M_p$ is
the mass in slit cell $p$, computed by Monte-Carlo integration. In
this case, we set $\sigma(h_n) = 0.005$ ($0.003)$ for the central slit
(integral field grid) cell $m_c$ to approximate realistic errors.

\subsection{Making Triaxial Models with NMAGIC}
\label{ssec:triModels}

\begin{figure}
\includegraphics[angle=-90.0,width=0.9\hsize]{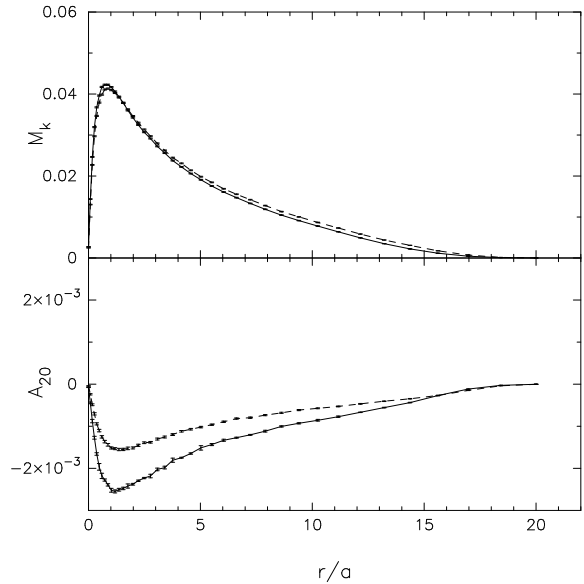}
\vskip0.5truecm
\caption[]{Target mass and $A_{20}$ profiles for the triaxial models.
  The solid line shows target T54 while the dashed line shows target
  T53.}
\label{fig:triAdiff}
\end{figure}

\begin{figure}
\includegraphics[angle=-90.0,width=\hsize]{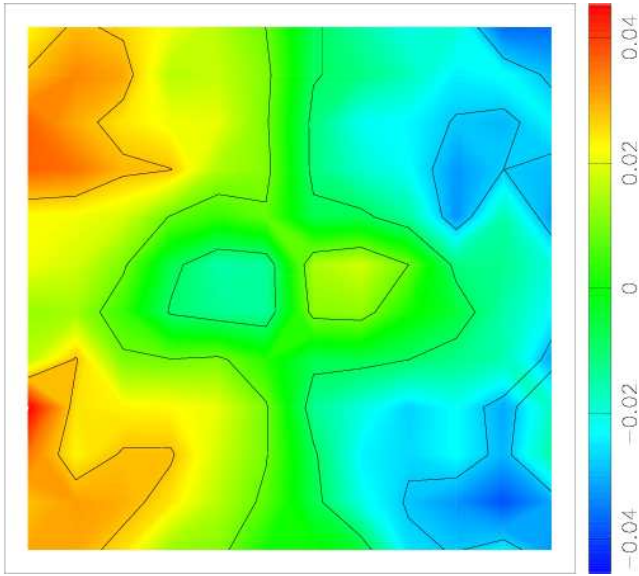}
\vskip0.5truecm
\caption[]{Line-of-sight velocity field of the rotating triaxial
  particle model (RT54K) as seen in the inertial frame. The co-rotation
  radius is $R_{\rm cor}\approx 10 R_{\rm eff}$. The FoV extends from
  $-R_{\rm eff}$ to $R_{\rm eff}$ along each direction. Its lower edge
  is parallel to the major axis, the line-of-sight is parallel to the
  intermediate axis.  Notice the counter-rotation near the center.}
\label{fig:triaxvr}
\end{figure}

\noindent
In the tests below we also explore triaxial Hernquist target models with 
stellar densities 
\begin{equation}
  \varrho (s) = \frac{M a}{2 \pi x_0 z_0 s(s+a)^3},
\label{eq:tridens}
\end{equation}
where $M$ is the total mass, $a$ the scale radius, and $s =
\sqrt{(x/x_0)^2+y^2+(z/z_0)^2}$.  Here $y$ is the longest axis, and
the parameters $x_0$ and $z_0$ are the axis ratios.  As before, we use
units with $M = a = 1$ and we define the effective radius with
reference to the spherical model, \ie $R_{\rm eff} \approx 1.815$.  We
generate two targets with different triaxialities, characterized by
the triaxiality parameter $T = (1-x_0^2)/(1-z_0^2)$
\citep{franx_etal91}. The more triaxial target, hereafter
T53, has $x_0 = 0.9$ and $z_0 = 0.8$ ($T = 0.53$) whereas the less
triaxial target, hereafter T54, has $x_0 = 0.85$ and $z_0 = 0.7$ ($T =
0.54$).  In both cases the target is observed along its intermediate
($x$-)axis.

Like our oblate target model, the triaxial models cannot be
represented by a DF based on the integrals of motion.  We therefore
construct them through particle realizations via a two step process.
Starting from a spherical Hernquist particle realization made from a
DF as before, we squeeze this along the x- and z-axes by factors $x_0$
and $z_0$, respectively, and compute the desired target density
observables $M_k$ and the higher order mass moments $A_{20,k}$,
$A_{22,k}$ up to $A_{60,k}$ using the same radial binning as in the
spherical and oblate targets.  $A_{lm}$ components with $l>6$ are
small and we omit them.  The squeezing is rigid, \ie without regard to
the internal motions.  We repeat this 30 times, squeezing the
spherical Hernquist model rigidly along random orientations to the
desired shapes.  From these 30 particle representations of the model
we compute the means and one $\sigma$ variations around the mean for
the $A_{lm,k}$.  The former are taken as target density observables,
the latter as their errors. The uncertainties on the radial mass in shells
profile are taken to be $\sigma(M_k)=\sqrt{M_k M_{\rm trun}/N}$ as
before.  Figure \ref{fig:triAdiff} shows the target mass and $A_{20}$
profiles as functions of radius for T54 (solid line) and T53 (dashed
line) as well as their uncertainties.

After this first step, which only gives target density observables, we
then use $\chi^2$M2M to evolve a spherical Hernquist model to generate
self-consistent triaxial particle realizations of T54 and T53. In
addition we generate a slowly tumbling version of T54 with corotation
radius $R_{cor}\approx 10 R_{\rm eff}$, by applying $\chi^2$M2M in the
appropriately rotating frame. The final models now have
self-consistent kinematics; in order to distinguish them from the
purely density targets we refer to them as models T53K and T54K for
the non-tumbling models and RT54K for the tumbling model.

These final self-consistent models T54K and RT54K can now be used as
targets in their own right, and we can compute (observer frame) target
kinematics $m_p h_{n,p}$ from them.  We compute the kinematics of both
T54K and RT54K on a $12\times 12$ grid extending from $-R_{\rm eff}$
to $R_{\rm eff}$.  For the uncertainties in the kinematic observables
we adopt $\sigma(m_p h_{n,p}) = \sigma(h_n) M_c\sqrt{M_p/M_c}$ with
$\sigma(h_n) = 0.005$ the error in $h_n$, $M_p$ the mass in grid cell
$p$, and $M_c$ the mass in the central grid cell. The $M_p$'s were
obtained directly from the particles.  The velocity field of the
target system RT54K in the observer's frame is shown in Figure
\ref{fig:triaxvr}.  This velocity field is characterized by disk-like
counter-rotation close to the mid-plane and near cylindrical rotation
away from the plane.  These kinematics for this slowly tumbling
triaxial model represent a valid dynamical model, but are unlikely to
be the unique dynamical solution for the model's density distribution.


\section{Tests of NMAGIC}
\label{sec:Tests}

In this section we will use the $\chi^2$M2M algorithm to solve some
modeling problems of increasing dimensionality and complexity,
starting with spherical systems and ending with rotating triaxial
models. The goal of these experiments is to investigate the
convergence of the code, the quality with which various data are
modeled, and the degree to which known properties of the target models
can be recovered from their simulated data. We will see how these
issues depend on the initial model, geometry, and amount of data
available. 

Table \ref{table:Tests} lists all the experiments that we have carried
out, including the target and the initial model identifications.  We
will refer to the final $\chi^2$M2M models by the prefix 'F' to the
test model name (e.g., FA for the final model of Test A).  Generally,
these final models are obtained in two steps.  First we use only the
target density observables in the $\chi^2$M2M algorithm, and once
these have converged, we add the kinematic observables. Finally, we
integrate all orbits for some time in the potential without
$\chi^2$M2M corrections to test whether equilibrium has been
reached. Unless mentioned otherwise, we use $1.8\times 10^6$ particles
and set the entropy parameter $\mu$ to a small ($\ll 1$) value; see
the discussion in Section 5.1.1.  In most experiments, the particle
distribution is evolved in the fixed target potential (this is
analogous to the Schwarzschild modelling approach), but we include one
test (model E) in which we also let the gravitational potential
evolve.

\begin{table}
\vbox{\hfil
\begin{tabular}{ccccccc}
  \hline
   \textsc{Test} & \textsc{ICs} & \textsc{Target} & $\varepsilon'$ &
   $\varepsilon''$ & $\mu$ \\
  \hline
  A  & RP    & SIH & $0.025$ & $6.32\;10^{-7}$ & $4.3\;10^{-4}$ \\
  A2 & RP    & SIH & $0.025$ & $1.32\;10^{-6}$ & $4.3\;10^{2}$ \\
  A3 & RP    & SIH & $0.050$ & $1.24\;10^{-6}$ & $4.3\;10^{-4}$ \\
  A4 & RP    & SIH & $0.100$ & $6.80\;10^{-7}$ & $4.3\;10^{-4}$ \\
  B  & SIH-2 & SIH & $0.025$ & $1.76\;10^{-6}$ & $4.3\;10^{-4}$ \\
%
  C & ORIH & O3I & $0.05$ & $3.94\;10^{-7}$ & $0$  & \\
  D & ORIH & O3I & $0.05$ & $3.94\;10^{-7}$ & $0$  & \\
%
  E & T53K & T54K  & $0.15$ & $5.06\;10^{-8}$ & $4.3\;10^2$ & \\
  F & T54K & RT54K & $0.15$ & $3.77\;10^{-8}$ & $4.3\;10^2$ & \\
  \hline
\end{tabular}
\hfil}
\caption{Tests of NMAGIC carried out in this paper, with model names
and parameters. For all models, we have used $\alpha=2.1 \varepsilon'$.}
\label{table:Tests}
\end{table}

\subsection{Spherical Models}
\label{ssec:ResSphMods}

\subsubsection{Initial model and time-evolution}

\begin{figure}
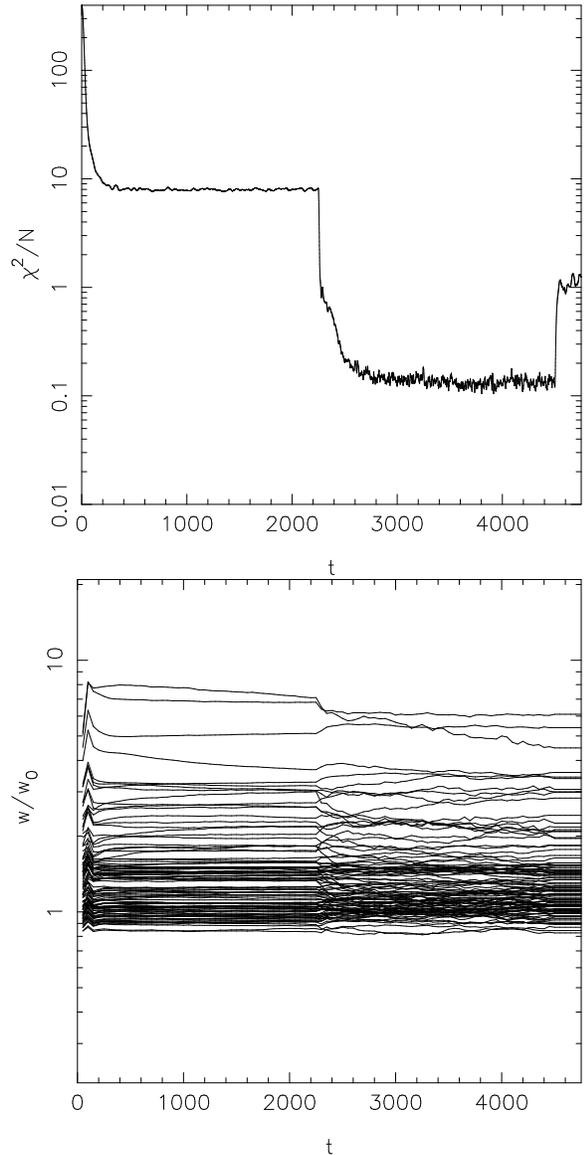

\includegraphics[angle=-90.0,width=0.90\hsize]{Fig8a_ChisqEvoln.ps}
\includegraphics[angle=-90.0,width=0.90\hsize]{Fig8b_wEvoln.ps}
\vskip0.2truecm
\caption[]{(a) Top: Time evolution of $\chi^2$ in test A.  (b) Bottom:
Time evolution of a set of 100 particle weights in test A. $w_0$ is
the initial weight of the particles; $w_0 = 1/N$.  The time-interval
plotted includes a first phase of density adjustment ($t \leq 2250$),
a second phase of density and kinematic adjustment ($2250 < t \leq
4500$), and a final phase of free evolution during which the weights
do not change ($t \geq 4500$).  Time is in units where the dynamical
time at the half-mass radius is 6.0, and the dynamical time at $r_{\rm
max}$ is 150.}
\label{fig:SIHc2}
\end{figure}

\begin{figure}
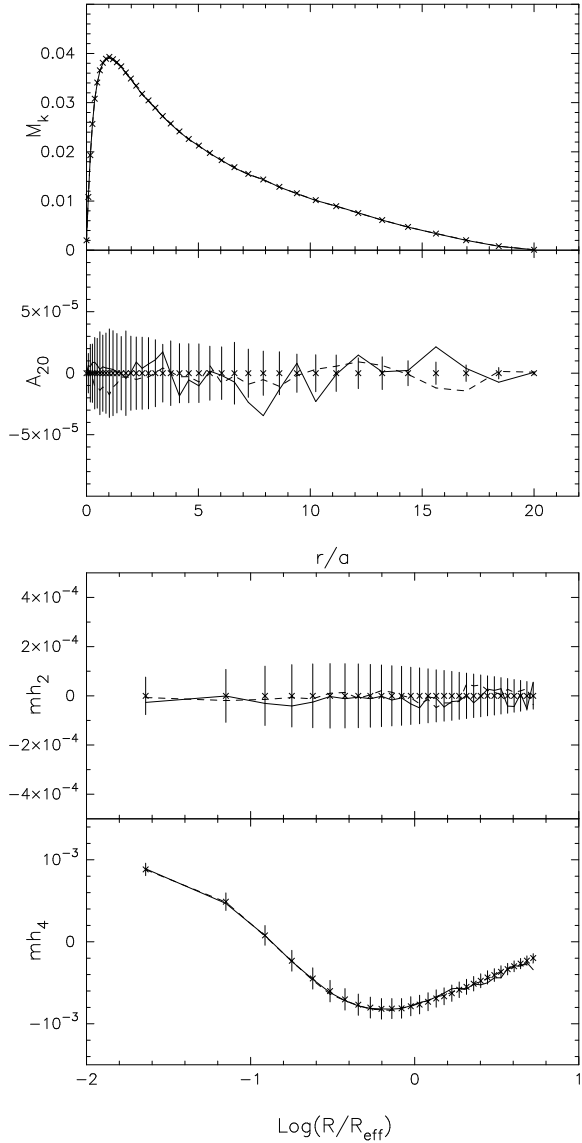

\includegraphics[angle=-90.0,width=0.90\hsize]{Fig9a_MA_FAB.ps}
\includegraphics[angle=-90.0,width=0.90\hsize]{Fig9b_mh_FAB.ps}
\vskip0.2truecm
\caption[]{(a) Top: Radial mass and $A_{20}$ profiles for the target
model SIH and the final models FA from Test A (solid line) and FB from
Test B (dashed line).  (b) Bottom: Kinematic profiles $m\,h_2$ and
$m\,h_4$, for the same models. -- In all panels, the data points with
errors correspond to the SIH target, the solid line to the final
particle model FA, and the dashed line to the final model FB. The
error bars in the target mass distribution are not shown as they are
smaller than the symbol sizes. The absolute errors shown decrease
outwards due to the mass weighting; the corresponding relative errors
increase outwards.}
\label{fig:SIHAlm}
\end{figure}

The aim of our first experiment, Test A, is to reproduce a spherical
isotropic Hernquist (SIH) model by a $1.8\times 10^6$ particle model.
We start by generating a Plummer model from its DF (\eg
\citealt{bin_tre87}), using the method described in
\citet{victor_sell00}.  The DF of the Plummer model is truncated at
$\Phi(r_{\rm max})$, with $r_{\rm max}=20$, and has a scale length
$b=1$ and unit total mass. We then relax these particles in the
analytic Hernquist potential, which is held fixed while the particle
orbits are integrated. We refer to the resulting particle distribution
as initial model RP (relaxed Plummer).

Then with $\chi^2$M2M we first adjust the density distribution of
model RP to that of the target SIH, using as target observables $M_k =
\sqrt{4 \pi} A_{00,k}^t$ (equation \ref{eqn:HernM}) and $A_{lm,k}^t=0$
for $1<l\le 6$, $0<m\le l$ (equation \ref{eqn:alm}) with Monte Carlo
errors estimated as described in Section \ref{ssec:sphModels}. After
convergence the even kinematic moment observables $M_k h_{2,k}$ and
$M_k h_{4,k}$ are added with errors given also in Section
\ref{ssec:sphModels}. Finally the system is integrated for some time
without applying the $\chi^2$M2M corrections.

The second experiment B is identical to A except that instead of model
RP we use a second Hernquist model SIH-2 as initial conditions for
NMAGIC. SIH-2 differs from the target model SIH in that its radial
scale length $a=1.4$ instead of $a=1$.

Figure \ref{fig:SIHc2}a shows the time evolution of $\chi^2/N_o$ of
the particle model A during and after the $\chi^2$M2M
evolution.  Throughout $N_o$ refers to all the observables, density
and kinematics, regardless of whether they are being used in the FOC
or not; thus $N_o$ is a constant.  The time evolution of a sample of
100 particle weights of the SIH particle model is presented in Figure
\ref{fig:SIHc2}b. From these figures one sees that the overall
$\chi^2/N_o$ decreases quickly at the beginning of both phases
(density adjustment only phase, and a density and kinematic observable
adjustment phase). However, particle weights keep evolving for
significantly longer time-scales. For this reason we integrate and
adjust particle weights in both phases for relatively long times,
about 15 dynamical times at $r_{\rm max}$.

\subsubsection{Convergence to the target observables for different
initial conditions}

\begin{figure}
\includegraphics[angle=-90.0,width=0.9\hsize]{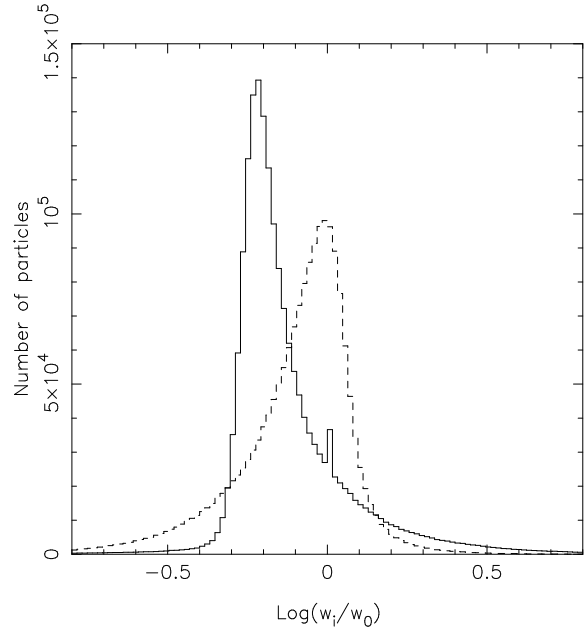}
\vskip0.2truecm
\caption[]{Histogram of the particle weights in the final FA model,
obtained from Plummer model initial conditions (solid line). The
dashed line shows the histogram of particle weights when spherical
Hernquist ICs with scale length $a=1.4$ were used (FB).  $w_0$ is the
initial weight of the particles; $w_0 = 1/N$ in all cases.}
\label{fig:SIHwhist}
\end{figure}

The fit of the final particle models FA and FB to the observables is
illustrated in Figure \ref{fig:SIHAlm}. The top panel shows the radial
mass and $A_{20}$ coefficient, whereas the bottom panel shows the
kinematic targets and final model observables for $m\,h_2$ in the
upper and $m\,h_4$ in the lower panel. As can also be seen
from Fig.~\ref{fig:SIHc2}, the final model fits the input data to
within $1\sigma$. The corresponding error bars are smaller than the
crosses in the top panel of Fig.~\ref{fig:SIHAlm}; see
Fig.~\ref{fig:triAdiff} for an example. The same is true for $A_{20}$
except when the target values are zero as in Fig.~\ref{fig:SIHAlm}.
Error bars for the mass observables are therefore not plotted in this
and subsequent similar figures.

All model observables in
Fig.~\ref{fig:SIHAlm} are temporally smoothed observables as in
equation~(\ref{eqn:myts}).  After some free evolution with $\chi^2$M2M
turned off both models fit the target data within the errors.  The
free evolution is necessary because $\chi^2$M2M pushes the model
towards a perfect fit to the observables, at the expense of
continually changing particle weights.  Deviations are largest in the
outer parts where orbital time-scales are longest.  Model FB, which
had an initial particle distribution closer to the target, is
generally smoother and fits the data better, but differences are
within the errors. NMAGIC achieved satisfactory models even from the
less favorable, cored Plummer initial conditions.

Figure \ref{fig:SIHwhist} compares the histogram of final particle
weights of the FA and FB models, all normalized by their initial
weight. Model FA has a significant tail towards high weights, and a
peak at correspondingly lower particle weight such that the mean
particle weight is the same as for the more symmetric weight
distribution of model FB.  On average, the weights of particles in
model FA had to change by more than those in model FB. We can
quantify this by defining an {\sl effective} particle number
$N_{\rm eff}$ characterizing mass fluctuations through
\begin{equation}
N_{\rm eff} \equiv N {\overline{w}^2 \over \overline{w^2}},
\label{eq:neff}
\end{equation}
where $\overline{w}$ and $\overline{w^2}$ are the mean and mean-square
particle weights. This reduces to $N$ for equal-mass particles, to one
when one particle dominates, and discards particles with near-zero
weights. For the final models FA and FB the effective numbers of
particles are $N_{\rm eff}= 5.7\times 10^5$ and $1.5\times 10^6$,
respectively, while for both models $N=1.8\times 10^6$.

\begin{figure}
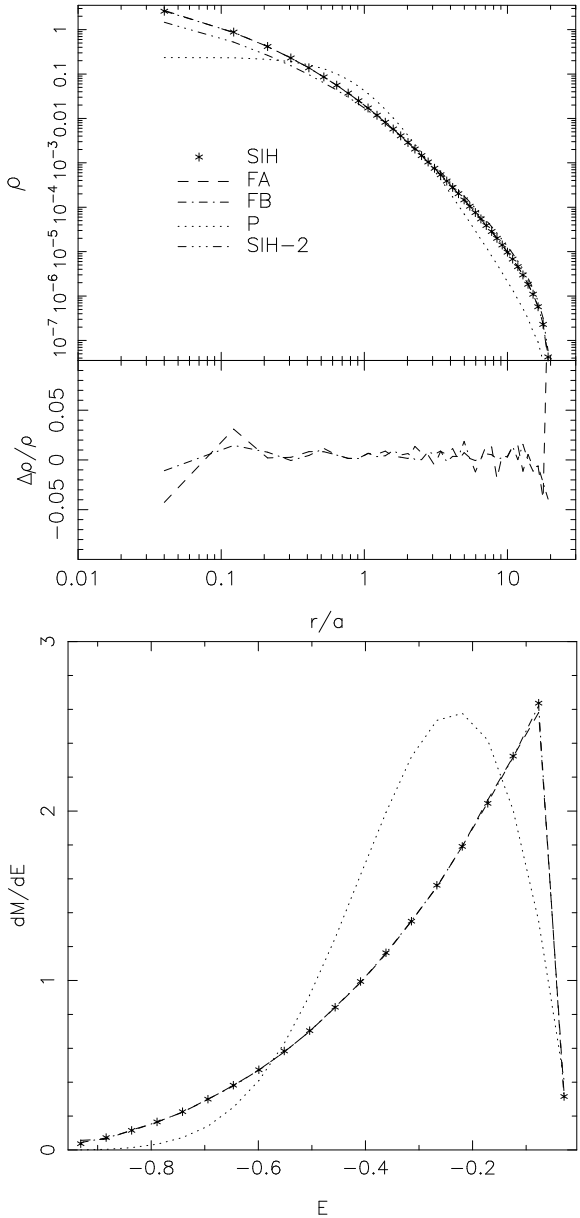

\includegraphics[angle=-90.0,width=0.90\hsize]{Fig11a_rho_FAB.ps}
\includegraphics[angle=-90.0,width=0.90\hsize]{Fig11b_df_FAB.ps}
\vskip0.2truecm
\caption[]{(a) Top: Radial density profiles in the spherical models.
Uppermost panel: Density profiles for the Hernquist target profile, SIH
(stars), the final models FA (dashed line) and FB (dash-dotted
line), and their respective initial condition models RP and SIH-2
(dotted and dash-triple-dotted lines). Middle panel: Relative deviation
from the target density $\uDelta\rho/\rho$, for the two models FA and
FB using the same line styles.  (b) Bottom: Differential energy
distributions. The truncated analytic Hernquist DF used for target
SIH is shown by the star symbols. The dashed line corresponds to the
final $\chi^2$M2M model FA, and the dotted line indicates the relaxed
Plummer initial conditions RP.}
\label{fig:dens}
\end{figure}

The origin of this difference between the two models can be seen from
Figure \ref{fig:dens}a, which plots the radial density profile of the
target SIH (stars), the initial models RP and SIH-2, and the
temporally smoothed final models.  We computed the densities using the
identical radial grid as was used for the mass targets.  The density
profile of the SIH target is well reproduced by the final particle
models FA and FB across more than a factor of 100 in radius.  The
largest relative deviation in the density $\delta\rho/\rho$ occurs at
small radii and never reaches more than $5\%$.  In this region, model
RP has few particles and the large relative error is due to Poisson
noise.  Model FB, which starts out closer to the target SIH fits
better in this region.

Model RP is clearly significantly less dense than SIH inside $r \simeq
0.3 a$; it has a core whereas the target profile is cuspy. Also, it
has a steeper outer density profile than the target model. To match
model RP to SIH therefore requires NMAGIC to increase the particle
masses both in the central regions and in the outer halo of the
model. This causes the high-weight tail in the distribution in
Fig.~\ref{fig:SIHwhist}, as we verified by inspecting the positions of
particles with $w_i > 2 w_0$.

Figure \ref{fig:dens}b presents the differential energy distributions.
The final particle model FA matches the analytic differential energy
distribution of the isotropic Hernquist model (equation
\ref{eqn:HernquistDF}) very well.

\begin{figure}
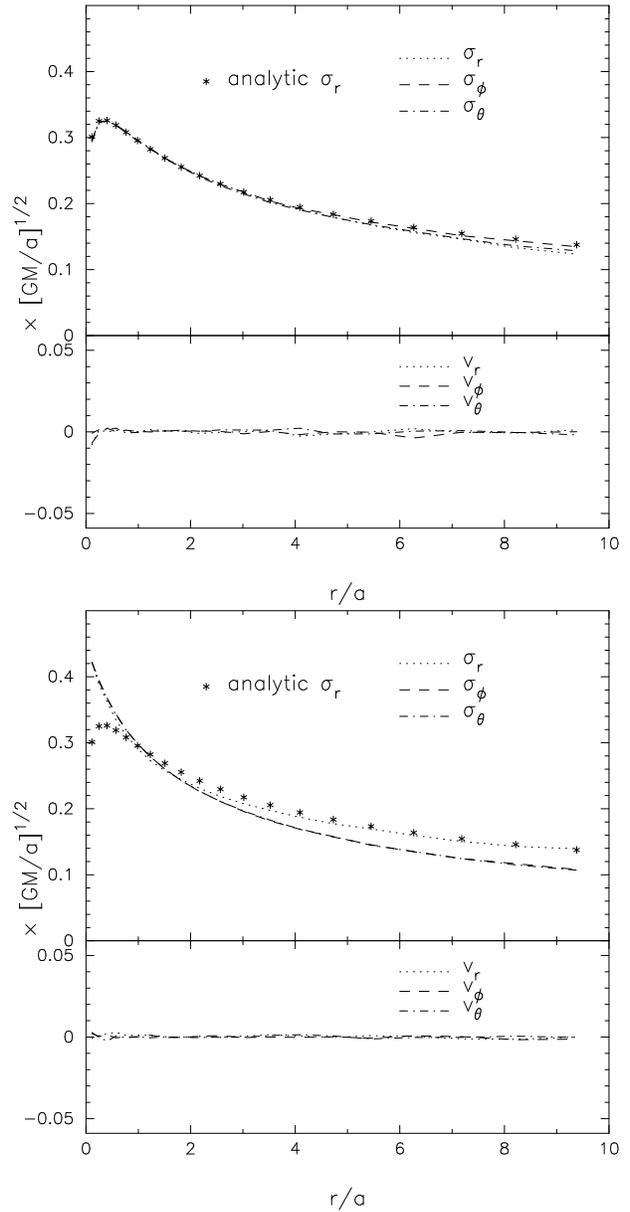

\includegraphics[angle=-90.0,width=0.96\hsize]{Fig12a_sig_FAB.ps}
\includegraphics[angle=-90.0,width=0.96\hsize]{Fig12b_sig_RP.ps}
\vskip0.2truecm
\caption[]{(a) (b) Top: Internal kinematics of the final model FA.
The upper panel show $\sigma_r$, $\sigma_{\varphi}$ and
$\sigma_{\theta}$, the lower panel the $v_r$, $v_{\varphi}$ and
$v_{\theta}$. The stars correspond to the analytic $\sigma_r$ from the
untruncated DF.  Model FA is very nearly isotropic and has negligible
rotation, despite starting from anisotropic initial conditions.
Bottom: Anisotropic internal kinematics of the initial
model RP. The dotted, dashed, and dash-dotted lines show $\sigma_r$,
$\sigma_{\varphi}$, and $\sigma_{\theta}$ of the RP particle
model. For comparison, the solid line corresponds to the analytic
$\sigma_r$ of the untruncated analytic DF of the SIH target model. }
\label{fig:RPdisp}
\end{figure}

As a final test, Figure \ref{fig:RPdisp}a shows the intrinsic
velocities (lower panel) and velocity dispersions (upper panel) of the
analytic, untruncated DF and the final $\chi^2$M2M model FA.  The
match to the target kinematics is good and model FA is nearly
isotropic, despite the fact that it has evolved from an initial RP
model that is moderately anisotropic. The anisotropy of the initial
model RP is shown in Figure \ref{fig:RPdisp}b which compares its
intrinsic velocity dispersions $\sigma_r$, $\sigma_{\varphi}$ and
$\sigma_{\theta}$ with the analytic $\sigma_r$ of the SIH target
model.  The residual anisotropy in model FA is caused by the relative
absence of radial orbits resulting from truncating the DF.

\subsubsection{Dependence on $\epsilon'$ and $\mu$}

\begin{figure}
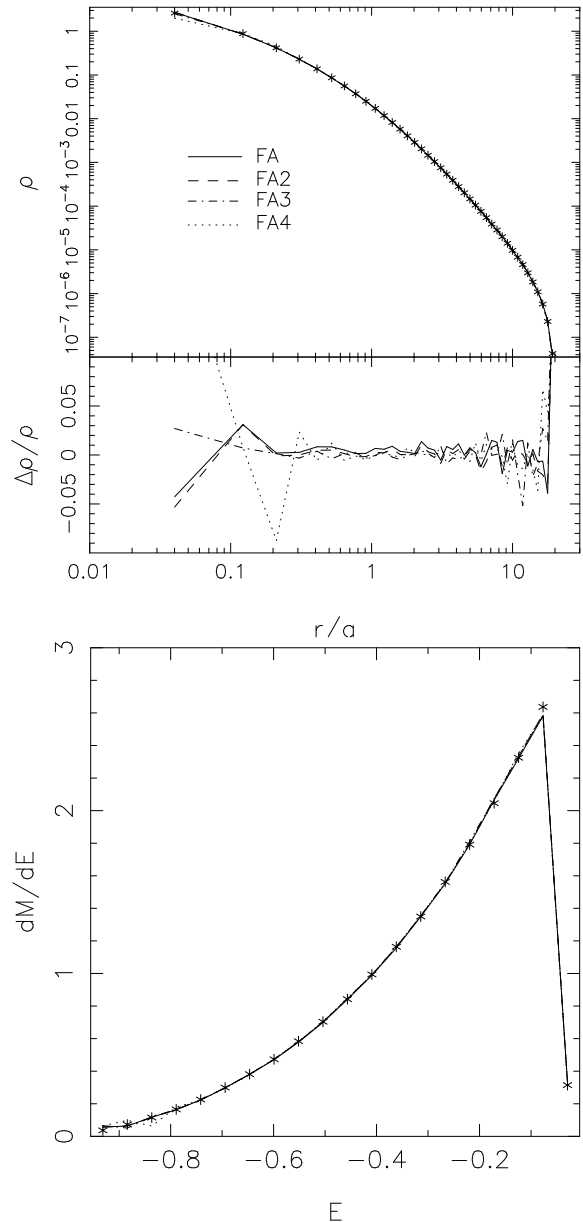

\includegraphics[angle=-90.0,width=0.90\hsize]{Fig13a_rho_EPS.ps}
\includegraphics[angle=-90.0,width=0.90\hsize]{Fig13b_df_EPS.ps}
\vskip0.2truecm
\caption[]{(a) Top: Radial density profiles for various spherical
models constructed for the Hernquist target profile, SIH.  Upper
panel: Density profiles for the target model (stars), the model FA
(dashed line) and several tests that differ from model FA by the
values of the parameters $\varepsilon'$ and $\mu$ (see Table
1). Middle panel: Relative deviation from the target density
$\uDelta\rho/\rho$, for the same models.  (b) Bottom: Differential
energy distributions. Stars: target model SIH. Lines: same models as
in top panel.}
\label{fig:densEPS}
\end{figure}

\begin{figure}
\includegraphics[angle=-90.0,width=0.9\hsize]{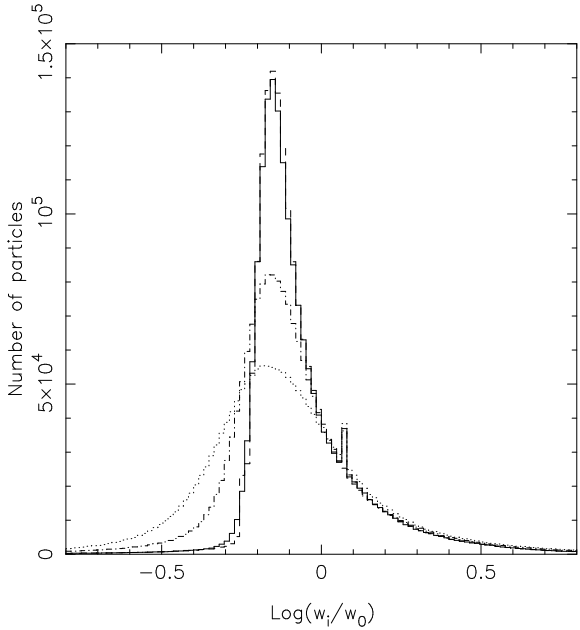}
\vskip0.2truecm
\caption[]{Histogram of the particle weights in the final FA model,
obtained from Plummer model initial conditions (solid line). The
other histograms show the particle weight distributions for models
FA2 (dashed), FA3(dash-dotted), and FA4 (dotted). $w_0$ is the
initial weight of the particles; $w_0 = 1/N$ in all cases.}
\label{fig:SIHwhistEPS}
\end{figure}

In the tests described so far, we have used $\epsilon'=0.025$ for the
correction steps in the FOC.  In general, small values of $\epsilon'$
result in a smooth evolution but slow convergence, whereas large
values of $\epsilon'$ change the global model too rapidly to attain a
properly phase-mixed stationary solution.  Thus generally we have
found $\epsilon'\apprla 0.1$ to give good results. This is illustrated
in Fig.~\ref{fig:densEPS}, which shows that test A converges to
essentially identical density distributions and differential energy
distributions for values of $0.025\le\epsilon'\le0.1$ (models FA, FA3,
FA4).  Only for the largest value $\epsilon'=0.1$ do we start seeing
small deviations in the density profile of more than a few percent
from the target model. Also, the effective particle number
[equation~(\ref{eq:neff})] decreases from $5.7\times 10^5$ through
$3.3\times 10^5$ to $1.0\times 10^5$ for models FA, FA3, and FA4,
respectively.  Thus we will generally use $\epsilon'<0.1$, but because
the speed of convergence also depends on the number and kind of
observables used for the corrections, we have sometimes also increased
$\epsilon'$ slightly. Figure~\ref{fig:SIHwhistEPS} shows the
distributions of particle weights for these models. They develop
larger wings for larger values of $\epsilon'$. Because particles
weights are then changed by larger amounts, the reshuffling is greater
until convergence is reached.

In models FA and FB, we have also set the entropy parameter $\mu$ to a
small ($\ll 1$) value, which allows the NMAGIC code to concentrate on
fitting the data.  (Note that, because the term $K_{ij}\,
\widetilde{\uDelta}_j / \sigma(Y_j) $ in the FOC is large, even
$\mu=1.0$ leads to only a small contribution of the entropy terms in
the FOC).  While the purpose of not setting $\mu$ to zero exactly
originally was to prevent overly large fluctuations in the particle
weights, in fact, a test with $\mu=0$ has given essentially identical
results to the ones reported. Fig.~\ref{fig:densEPS} shows that also
for model FA2 with $10^6$ times larger entropy parameter than in model
FA, the target density and differential energy distribution are fitted
equally well as before.  Generally, the best value to use for the
entropy depends on the initial model, the data to be fitted, and the
intrinsic structure of the target, and it must be determined
separately for each application. A more systematic investigation of
the effect of the entropy term is therefore deferred to a future paper
in which we will use $\chi^2$M2M to model and determine mass-to-light
ratio, anisotropy, etc., for a real galaxy.

\subsection{Oblate Models}
\label{ssec:ResObMods}

The task we set the algorithm here is a difficult one: starting from a
non-rotating system, we see whether we can recover the maximally
rotating three-integral model described in
Section~\ref{ssec:obModels}, in which the weights of all
counter-rotating particles should be zero. We perform two such
experiments, one using slit data as kinematic targets (Test C), the
second using integral field kinematic targets (Test D).  As in the
spherical experiments, we keep the potential fixed while evolving the
system with $\chi^2$M2M in runs C and D.

Both experiments start from an initial model which is constructed by
relaxing a spherical Hernquist particle model consisting of $5\times
10^5$ particles in the oblate potential.  As in experiments A and B,
we then apply $\chi^2$M2M in 2 steps, first for the density alone, and
when this has converged, for both the density and kinematics.  The
density part of the runs is identical for experiments C and D.

\begin{figure}
\includegraphics[angle=-90.0,width=0.9\hsize]{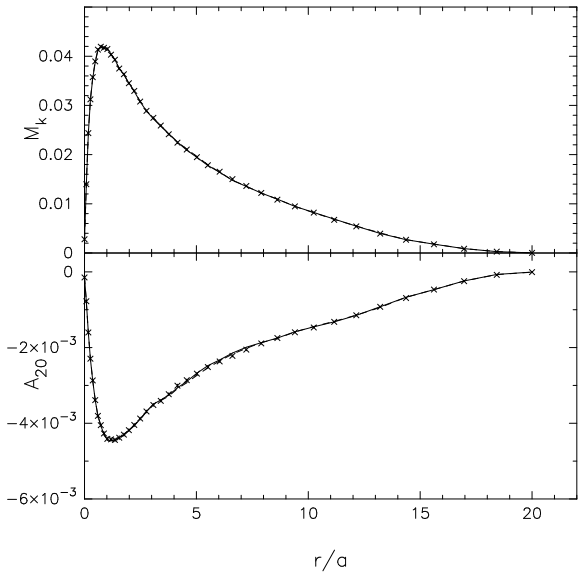}
\vskip0.5truecm
\caption[]{The mass and mass $A_{20}$ profiles for the oblate
 models. The data points show the target and the lines shows
 the converged models FC (dashed) and FD (full).}
\label{fig:AlmISO}
\end{figure}

\begin{figure}
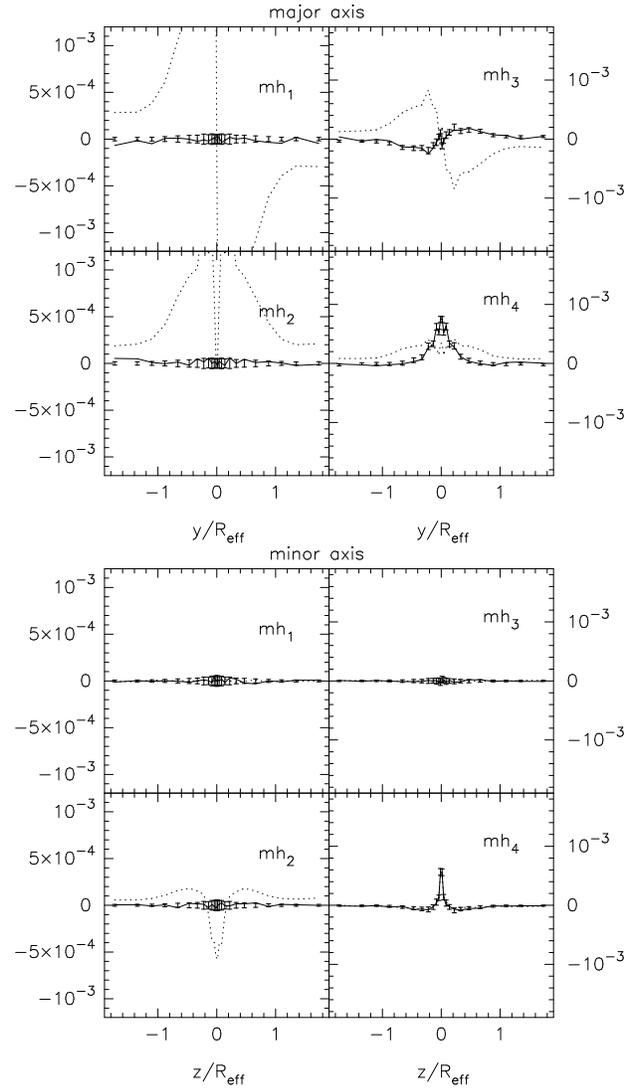

\includegraphics[angle=-90.,width=0.96\hsize]{Fig16a_mh_FCD_major.ps}
\includegraphics[angle=-90.,width=0.96\hsize]{Fig16b_mh_FCD_minor.ps}
\caption{Mass-weighted higher order moments along the major and minor
 axes for the slit-reconstructed oblate model FC.  The target observables
 are shown as error bars, whereas the model observables for model FC
 are indicated by the dashed lines, respectively.  Kinematics along
 the major axis are shown on the left and those along the minor axis
 on the right.
\label{fig:oblmajmins}}
\end{figure}

\begin{figure}
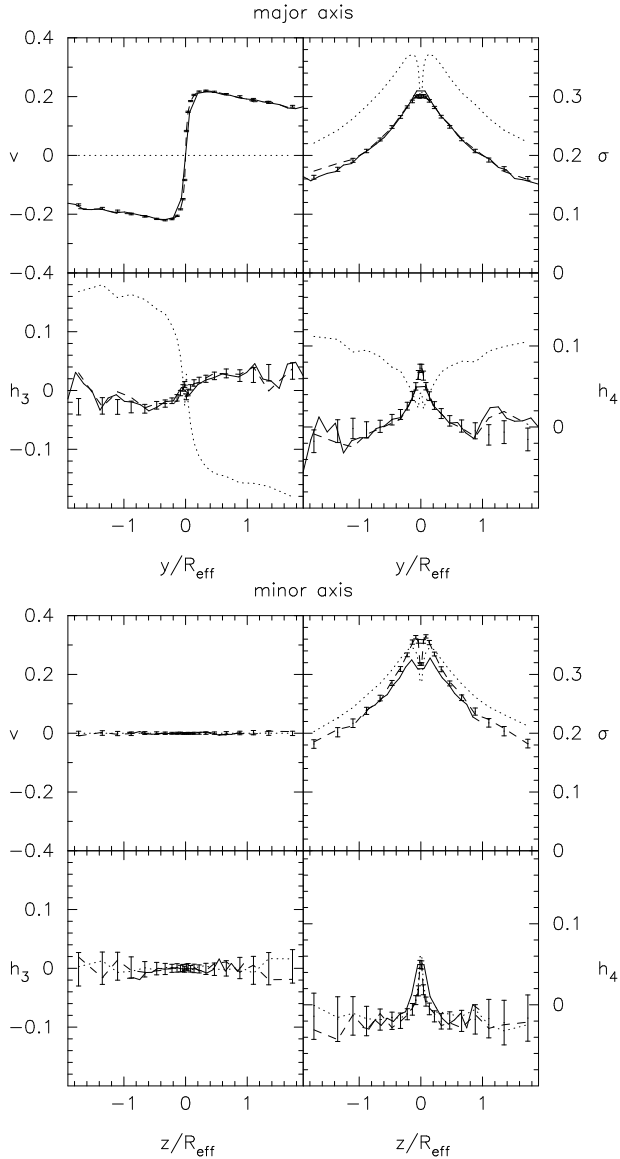

\includegraphics[angle=-90.,width=0.96\hsize]{Fig17a_vs_FCD_major.ps}
\includegraphics[angle=-90.,width=0.96\hsize]{Fig17b_vs_FCD_minor.ps}
\caption{Gaussian best fit velocity (top left), velocity dispersion
  (top right), Gauss-Hermite moments $h_3$ (bottom left) and $h_4$
  (bottom right) along the major axis (left) and minor axis (right),
  for the models with slit data targets (dashed line), integral field
  kinematic targets (full), and the initial model (dotted). The error
  bars show the target kinematics.}
\label{fig:oblmajfit}
\end{figure}

\begin{figure}
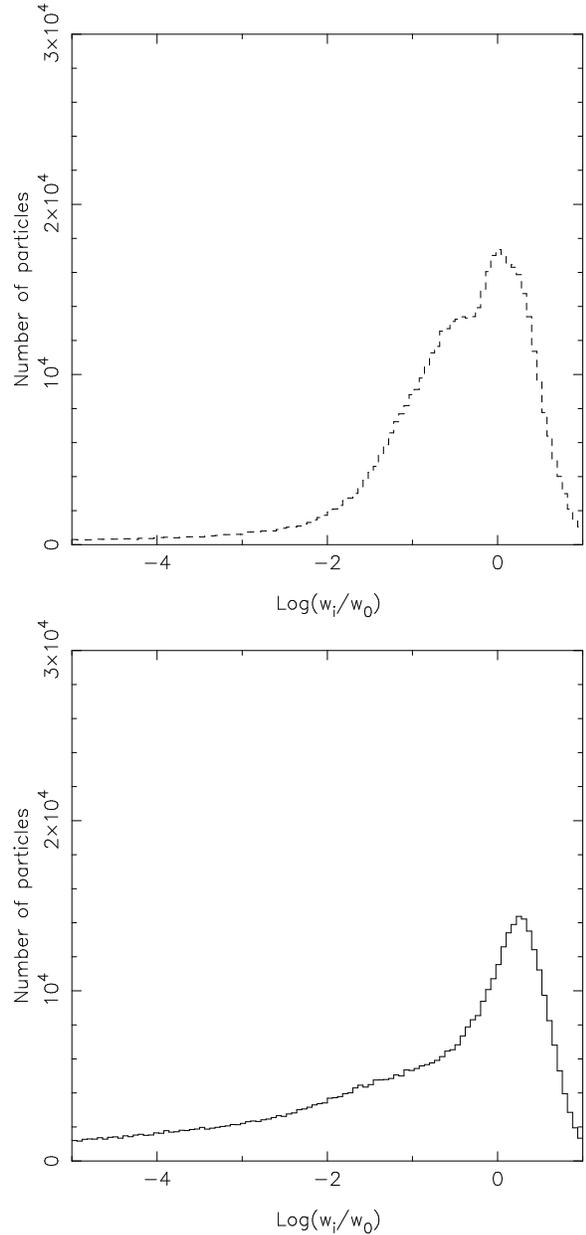

\includegraphics[angle=-90.,width=0.9\hsize]{Fig18a_wdist_FC.ps}
\includegraphics[angle=-90.,width=0.9\hsize]{Fig18b_wdist_FD.ps}
\vskip0.5truecm
\caption[]{Distribution of particle weights in the final models FC
(top) and FD (bottom).}
\label{fig:oblateweights}
\end{figure}

\begin{figure*}
\centerline{
\includegraphics[angle=-90.,width=0.5\hsize]{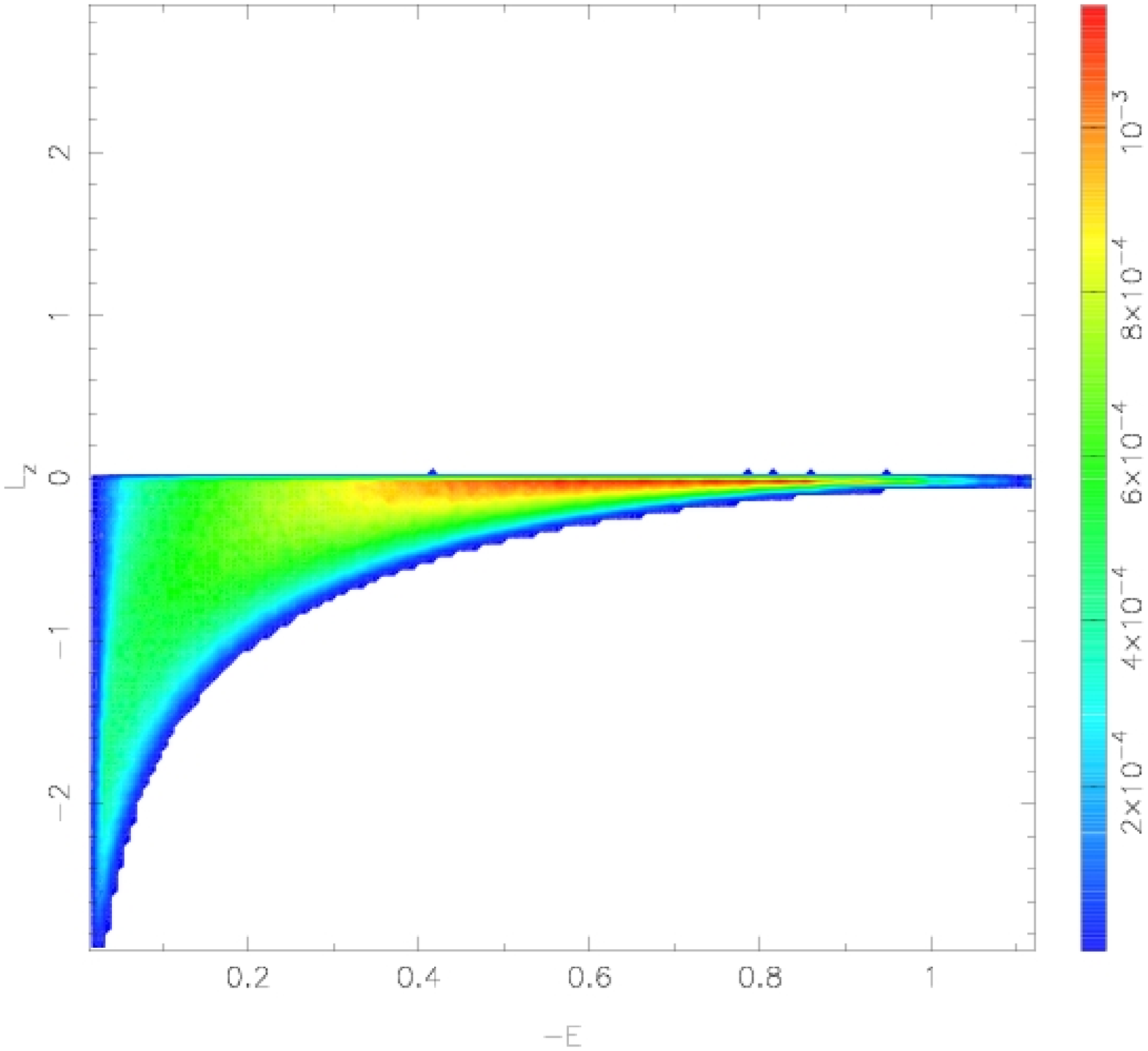}
\includegraphics[angle=-90.,width=0.5\hsize]{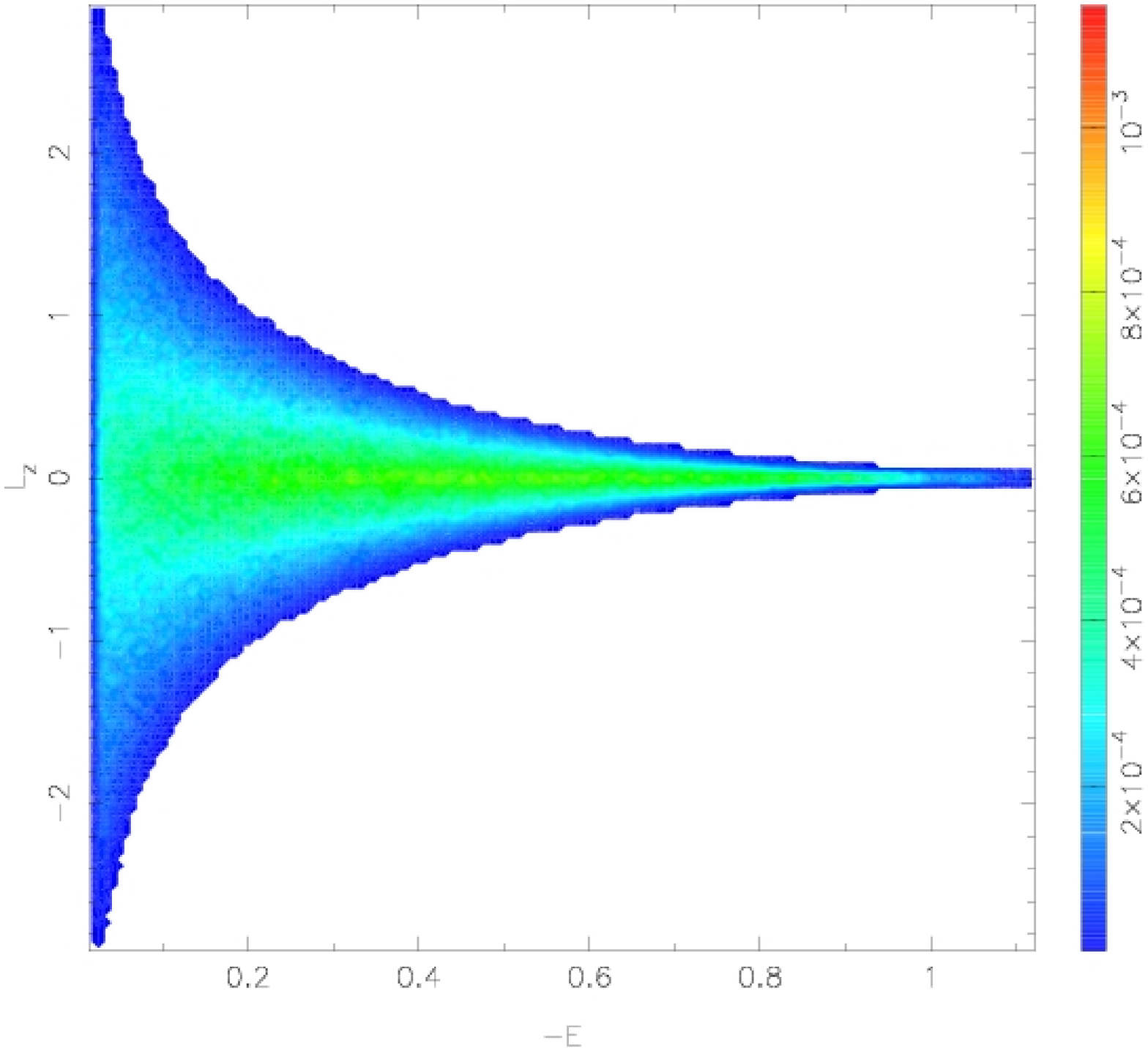}
}
\centerline{
\includegraphics[angle=-90.,width=0.5\hsize]{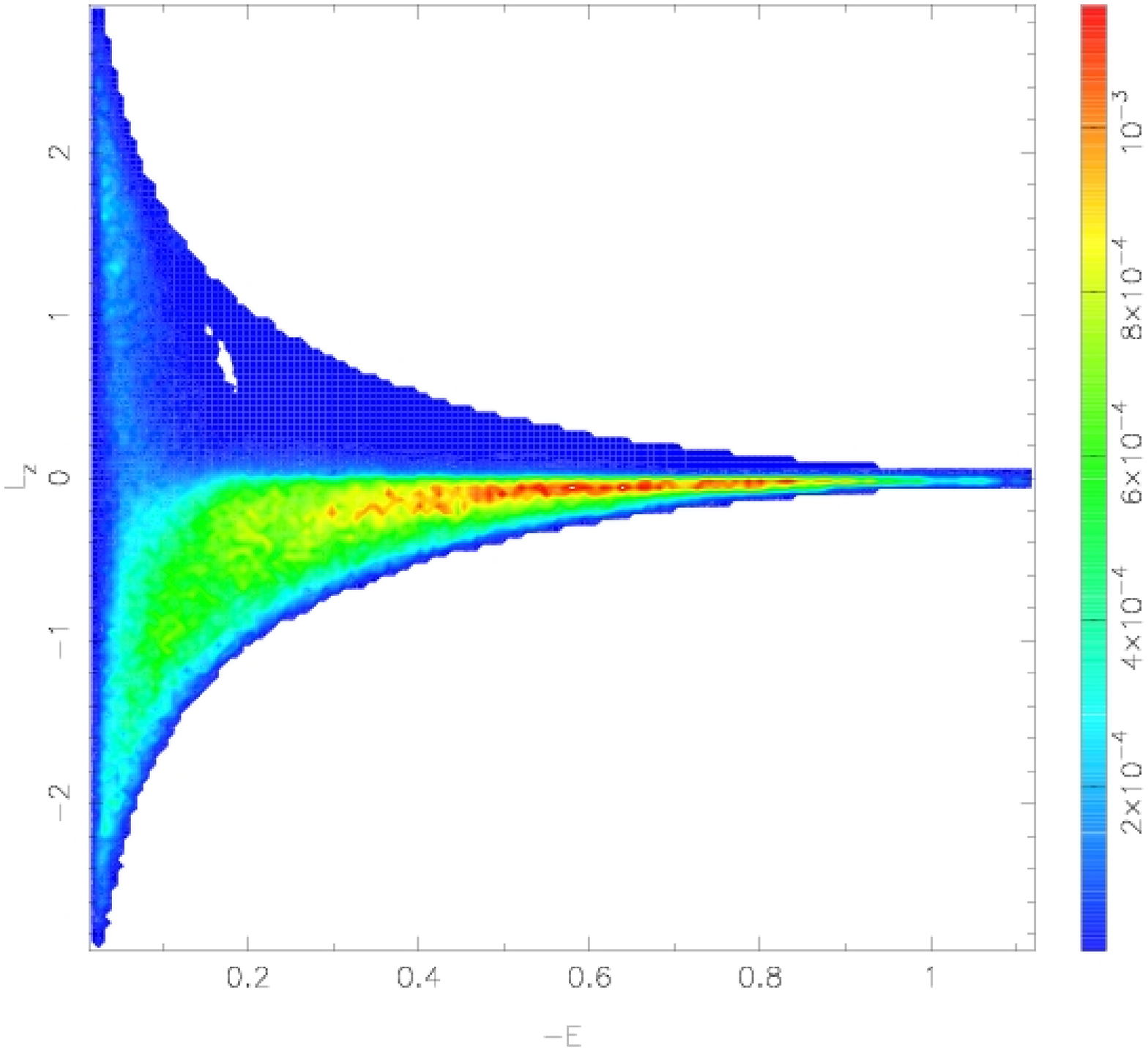}
\includegraphics[angle=-90.,width=0.5\hsize]{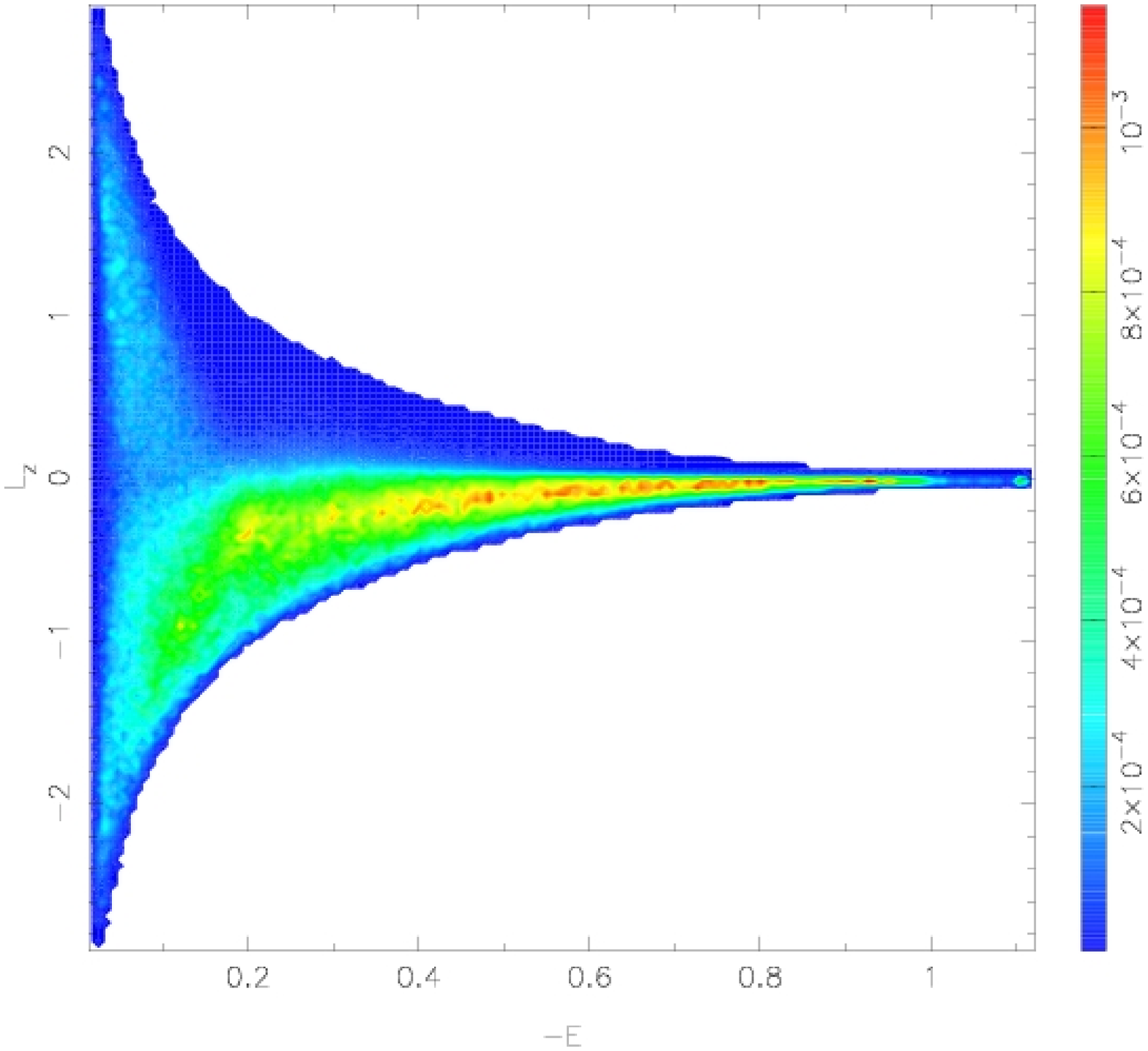}
}
\vskip0.5truecm
\caption[]{Particle weight distributions projected onto the
($E,L_z$) plane, for the maximally rotating three-integral target 
(top left), the initial relaxed isotropic Hernquist model (top right),
and the two models reconstructed from density and slit kinematic
targets (FC, bottom right) and from density and integral field
kinematics (bottom left).}
\label{fig:oblateELz}
\end{figure*}

Figure \ref{fig:AlmISO} plots the mass and $A_{20}$ radial profiles of
the target (error bars) and the final $\chi^2$M2M models FC
and FD. As in the spherical tests, the target density distribution
is very well fitted by the $\chi^2$M2M models.

The mass-weighted kinematics along the major and minor axes of model
FC are shown in Figure \ref{fig:oblmajmins}, while Figure
\ref{fig:oblmajfit} show the as-observed
kinematics of both models. The latter are calculated by dividing the
mass-weighted moments by the mass in the slit resp.\ grid cell, and
using the relations $v=v_{\rm targ} - \sqrt{2} \sigma_{\rm targ} h_1$
and $\sigma = \sigma_{\rm targ} - \sqrt{2} \sigma_{\rm targ} h_2$
(\eg, \citealt{rix_etal97}). All kinematic quantities for the
reconstructed models are shown $\Delta t=500$ ($\apprga 3$ dynamical
times at $r_{\rm max}$) after switching off the $\chi^2$M2M
corrections. The fits are generally excellent except for the higher
order moments near the boundaries of the kinematic fit regions, where
counter-rotating particles with high energies still make significant
contributions, because their weights have not yet been sufficiently
reduced.

Figure~\ref{fig:oblateweights} showing the weight distributions for
both models FC and FD clearly illustrates the stronger constraints
placed on the model by the integral field data. In both models, the
NMAGIC code works at reducing the weights of the counter-rotating
particles, but has clearly gone a lot further in model FD.

Finally, in Figure~\ref{fig:oblateELz} we show the distribution
of weights in the ($E,L_z$) plane for the target model, initial
relaxed model, and the two models FC and FD. The success of the
$\chi^2$M2M method in removing the counter-rotating particles 
amply present in the initial model is apparent, particularly for
model FD. Of course, in applications aimed at obtaining a best-fit
representation of some galaxy kinematic data it would have been
smart to start the iterations from an initial model that is
better adapted to the problem at hand.

\subsection{Triaxial Models}
\label{ssec:ResTriMods}

\subsubsection{ Evolving the potential self-consistently}

We illustrate NMAGIC's capabilities with two very different triaxial
model experiments.  In run E, we start with the self-consistent model
T53K as initial conditions and use NMAGIC to converge to target T54K.
With this model, we test the full capabilities of
$\chi^2$M2M, which make this technique more general than
Schwarzschild's method: in model E, we solve for the potential as the
system evolves and follow the model in its self-consistent potential
throughout, akin to an $N$-body experiment. For this purpose we use
the spherical harmonic potential solver described in Section
\ref{sec:chi2m2m} above and update the potential after every 25
$\chi^2$M2M steps.

\begin{figure}
\includegraphics[angle=-90.,width=0.9\hsize]{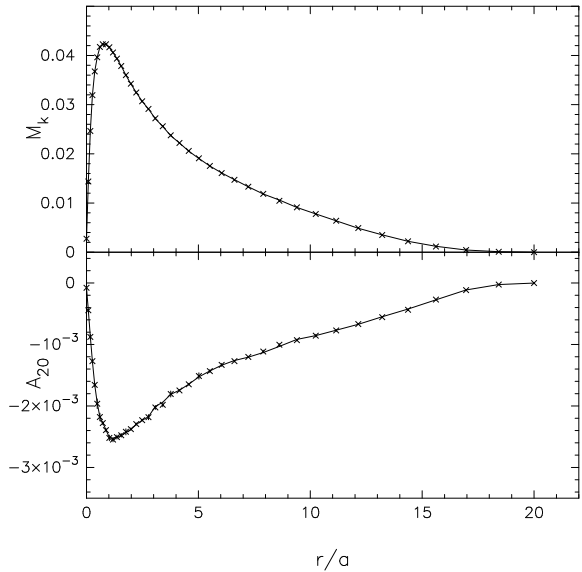}
\caption[]{Mass and $A_{20}$ profiles for experiment E. The dots show
  the target T54K and the solid lines show FE.}
\label{fig:triAlm}
\end{figure}

\begin{figure}
\includegraphics[angle=0.,width=\hsize]{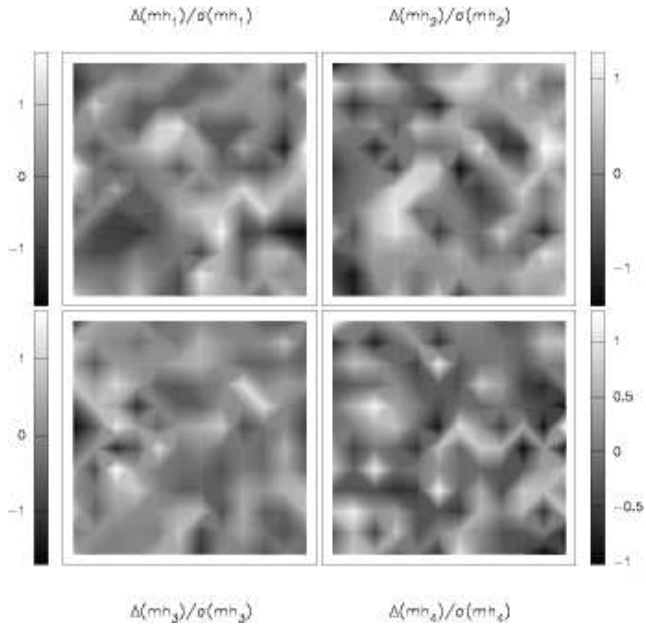}
\caption[]{The difference between kinematics in model T54K and model
FE.  The observables of FE are the temporally smoothed mass-weighted
moments while those of T54K are not temporally smoothed.
The differences have been divided by the corresponding assumed
errors. The FoV extends from $-R_{\rm eff}$ to $R_{\rm eff}$ along
each direction.}
\label{fig:tridiffnr}
\end{figure}

The resulting final model FE gives an excellent match to the density
of the target model T54K, as is apparent from comparing the $M_k$ and
$A_{20}$ profiles in Figure \ref{fig:triAlm}.
Figure \ref{fig:tridiffnr} shows the kinematics within $R_{\rm
eff}$ of the models T54K and FE.  All mass-weighted kinematic observables
$m\,h_{1}$,..., $m\,h_{4}$ of the final model match the target
observables at better than one $\sigma$ over almost the entire FoV,
except for a few isolated regions reaching two $\sigma$.  The random
location of these deviations imply that they are due only to Poisson
noise in the target model, the observables of which have not been
temporally smoothed.

\subsubsection{Rotating vs.\ non-rotating models}

Test F is an interesting experiment in different ways.  Starting from
T54K, we use NMAGIC to attempt to converge to the observables of the
tumbling target model RT54K, with a triaxial model which does not
tumble but remains stationary relative to the observer.  Thus this
experiment explores whether it is possible to identify a kinematic
signature of slow figure rotation in elliptical galaxies.  Since the
initial conditions possess neither rotation nor internal net stellar
streaming, if this model fails to converge it may well be because the
problem admits no solution.  Because of this, test F is interesting in
its own right, apart from as a validation of NMAGIC.

\begin{figure}
\includegraphics[angle=0.,width=\hsize]{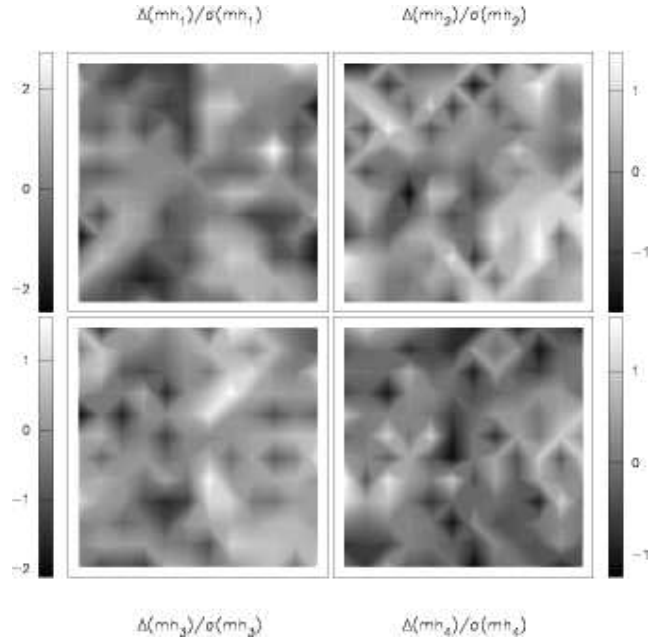}
\caption[]{The difference between kinematics in target RT54K and model
FF.  The observables are the mass-weighted higher order moments, and
have been divided by the corresponding assumed errors. The kinematics
of RT54K are instantaneous but those of FF are time-averaged.  The FoV
extends from $-R_{\rm eff}$ to $R_{\rm eff}$ along each direction.}
\label{fig:tridiffr}
\end{figure}

\begin{figure}
\includegraphics[angle=-90.,width=\hsize]{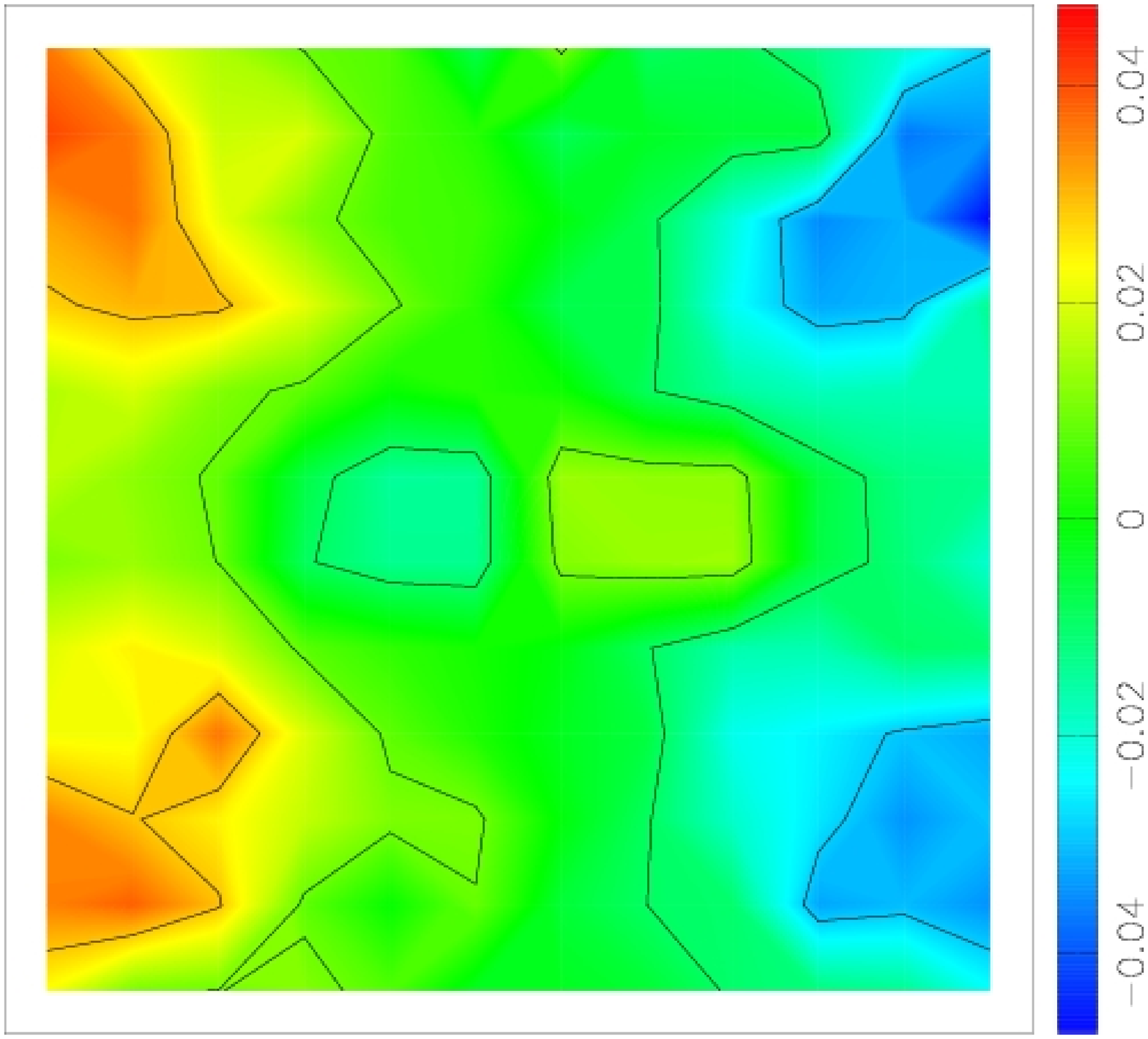}
\includegraphics[angle=-90.,width=\hsize]{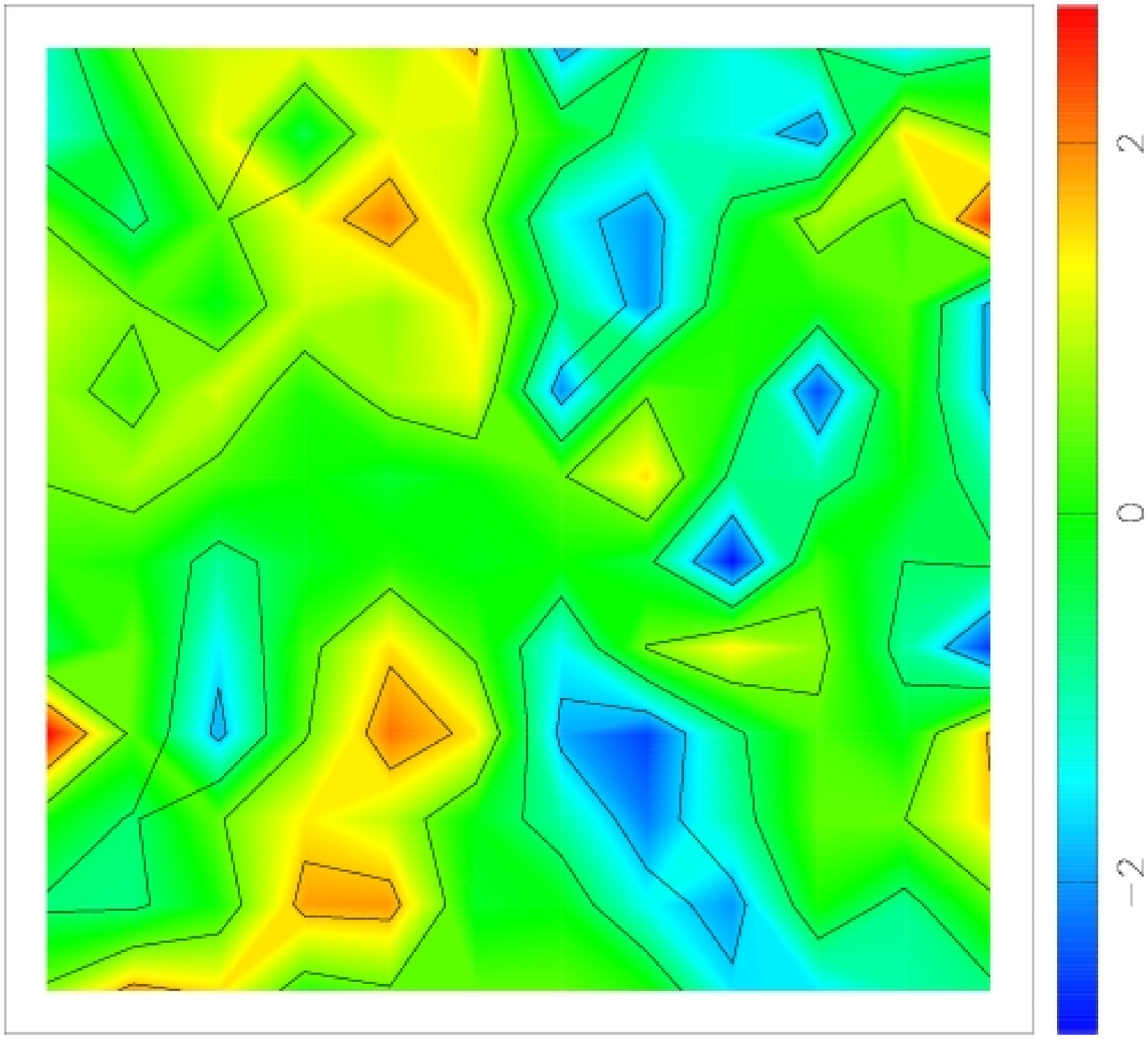}
\caption[]{Left: Line-of-sight velocity field of the final
non-tumbling triaxial particle model FF.  Right: Difference of the
line-of-sight velocity fields between the non-tumbling triaxial
particle model FF and the tumbling target galaxy RT54K divided by the
errors as described in the text.  We assume an error in the mean
velocity $\sigma(v_j) = \sqrt{2}\sigma(h_1)\sqrt{m_c/m_j}\sigma_j$,
where we assumed $\sigma(h_1) = -\sqrt{1/2}\sigma(v)/\sigma$
\citep{rix_etal97}.  In both panels the FoV extends from $-R_{\rm
eff}$ to $R_{\rm eff}$ along each direction.  }
\label{fig:triaxvnr}
\end{figure}

In fact, NMAGIC was able to converge the mass-weighted kinematic
moments to within about one $\sigma$ of their target values; however,
the residuals maps (Figure \ref{fig:tridiffr}) show spatially
correlated residuals in $m h_1$.  When we compare the global velocity
field of model FF with that of RT54K we find that the degree of
cylindrical rotation around the tumbling axis ($z$-axis) is higher in
RT54K than it is in model FF (Figure \ref{fig:triaxvnr}).  Near the
mid-plane, instead, the velocity field of both models is very similar,
including the counter-rotation seen near the center.  We can explore
whether the residual differences are due to having assumed too large
errors in the mass-weighted moments by decreasing the errors by a
factor of five. The corresponding final model looks very similar to
model FF but now with reduced $\chi^2 > 4$.  Thus the difference is
likely intrinsic and can be used to recognize a tumbling galaxy.  A
more complete analysis of this problem will be undertaken elsewhere
(De Lorenzi \etal in progress).

\section{Conclusions}
\label{sec:Conclusions}

We have presented a made-to-measure algorithm for constructing
particle models of stellar systems from observational data, building
on the made-to-measure method of \citet[ST96]{ST96}.
An important element of our new method is the use of the standard
$\chi^2$ merit function at the heart of the algorithm, in place of the
relative error used by ST96.  The improved algorithm, which we label
$\chi^2$M2M, allows us to assess the quality of a model for a set of
target data directly, using a statistically well-defined quantity
($\chi^2$).  Moreover, this quantity is well-defined and finite also
when a target observable takes on zero values.

This property has enabled us to incorporate kinematic observables
including higher-order Gauss-Hermite moments into the force-of-change
equation. Kinematic and density (or surface density) observables can
then be used simultaneously to correct the particle weights.  The
price of changing to $\chi^2$M2M from the original formulation is that
the kernels which project the particle weights and phase-space
coordinates into model observables cannot themselves depend on the
particle weights.  In general this is quite natural for (volume or
surface) density observables.  For the kinematics this means that we
need to use mass-weighted kinematic observables.  Nonetheless, this is
not a significant limitation.

We have implemented the $\chi^2$M2M method in a fast, parallel code,
NMAGIC.  This code also incorporates an optional but fast potential
solver, allowing the potential to vary along with the model density.
Its implementation of the $\chi^2$M2M algorithm is highly efficient,
with a sequential fraction of only $\sim 1\%$.  This has allowed us to
build various models with large numbers of particles and based on many
observables, and to run them for $\sim 10^6$ steps.

Then we have carried out a number of tests to illustrate the
capabilities and performance of NMAGIC, employing spherical, oblate
and triaxial target models.  The geometric flexibility by itself is
one of the main strengths of the method -- no symmetry assumptions
need to be made.

In the spherical experiments NMAGIC converged to the correct isotropic
model from anisotropic initial conditions, demonstrating that a unique
solution, if present, can be recovered.  Both the truncated
distribution function and the intrinsic velocity dispersions were
recovered correctly.  Two initial models with different density
distributions were used in these experiments. While both converged to
the final isotropic model, that with density closer to the density of
the final model had smaller final deviations from the target
observables, and a narrower distribution of weights. In both
experiments, the observables (density and integral field-like
kinematics) each converged in a few dynamical times at the outer
boundary $t_{d,o}$, whereas the particle weights kept evolving for
significantly longer, $\sim 10 t_{d,o}$.

In the oblate experiments we gave the algorithm a difficult problem to
solve. The target system was a maximally rotating three-integral model
in which the weights of all counter-rotating particles were zero. Using
density observables and either slit or integral field kinematics,
NMAGIC was asked to recover this maximally rotating model starting
from an isotropic spherical system relaxed in the oblate
potential. After about 100'000 correction steps, particle weights on
the counter-rotating side were reduced by a factor of $\sim 50$, the
distribution of weights approached that of the target, and a good fit
to the kinematic constraint data was achieved.  Only near the boundary
of the kinematic data did particles on orbits further out, whose
weights had not yet converged, still cause some deviations from the
target kinematics. These experiments also clearly showed the advantage
of integral field data over slit data for constraining the model.

Our triaxial experiments showed that it is possible to start from one
triaxial model and converge to another.  We anticipate that this
ability will be very useful in constructing models for the triaxial
elliptical galaxies with which nature confronts us.  One of these
triaxial experiments included a potential update step every 25
$\chi^2$M2M steps, demonstrating that including an evolving potential
is also practical.

In the final experiment, we first generated a particle model of a
slowly tumbling triaxial system to use as a target.  We then matched
its volume density and line-of-sight kinematics with a stationary
model.  We showed that the mass-weighted kinematic moments of the
figure rotating system was fitted to within one $\sigma$ by the
non-rotating system out to $R_{\rm eff}$.  However the residuals in
the first order kinematic moment are correlated, which gives a clear
signature of tumbling which the non-tumbling model is not able to
match, even when the assumed errors are decreased by a significant
factor.  We thus conclude that, at least for this triaxial system, it
is possible to distinguish between internal stellar streaming and
pattern rotation within $R_{\rm eff}$ provided a full velocity field
is available.  A more complete study of this problem will be presented
elsewhere.

This experiment also demonstrates the usefulness of the $\chi^2$M2M
algorithm for modeling mock (rather than real) galaxies in order to
learn about their dynamics.  We note that such an experiment would not
have been practical with standard $N$-body simulations.

Compared to the Schwarzschild method, the main advantages of the
$\chi^2$M2M algorithm as implemented in NMAGIC are that (i) stellar
systems without symmetry restrictions can be handled relatively
easily, (ii) it avoids complicated procedures for sampling, binning,
and storing orbits, and (iii) the potential can be evolved
self-consistently if needed. In the examples given, a simple isotropic
spherical model was evolved into a suitable initial model, which
contained the required wide range of orbital shapes.  Every
$\chi^2$M2M model corresponds to a new set-up of a complete orbit
library in the Schwarzschild method; so in problems where the same
orbit library can be reused, Schwarzschild's method will be
faster. However, NMAGIC is highly parallel, so suites of models with
$\sim 10^6$ particles are feasible on a PC cluster.

There is clearly room for improving the current implementation of
the $\chi^2$M2M algorithm, and there is a need to study carefully
the parameters that enter the algorithm, such as magnitude and
frequency of the correction steps, entropy, etc., which we will
address in future work.

However, the different applications presented in this paper show that
the $\chi^2$M2M algorithm is practical, reliable and can be applied to
various dynamically relaxed systems.  High quality dynamical models of
galaxies can be achieved which match targets to $\sim 1\sigma $ for
plausible uncertainties in the observables, and without symmetry
restrictions. We conclude that $\chi^2$M2M holds great promise for
unraveling the nature of galaxies.

\bigskip
\noindent

\section*{Acknowledgments}

We thank Scott Tremaine for comments on the manuscript and an
anonymous referee for his careful reading of the paper.  F.d.L.,
O.G. and N.S. are grateful to the Swiss National Science Foundation
for support under grant 200020-101766.  V.P.D. is supported by a
Brooks Prize Fellowship at the University of Washington and receives
partial support from NSF ITR grant PHY-0205413.  V.P.D. thanks the
Astronomisches Institut der Universit\"at Basel and the
Max-Planck-Institut f\"ur Ex. Physik for their hospitality during
parts of this project.
 
\bibliography{ms}

\label{lastpage}

\end{document}